# Rotation Curves in z~1-2 Star-Forming Disks: Evidence for Cored Dark Matter Distributions

R. Genzel[1,2], S.H. Price[1], H. Übler[1], N.M. Förster Schreiber[1], T.T. Shimizu[1], L.J. Tacconi[1], R. Bender[1,3], A. Burkert[3], A. Contursi[1,4], R. Coogan[1], R.L. Davies[1], R.I. Davies[1], A. Dekel[5], R. Herrera-Camus[6], M.-J. Lee[1], D. Lutz[1], T. Naab[7], R. Neri[4], A. Nestor[8], A. Renzini[9], R. Saglia[1,3] , K. Schuster[4], A. Sternberg[1,8,10], E.Wisnioski[11,12] & S. Wuyts[13]

*[1]Max-Planck-Institut für Extraterrestrische Physik (MPE), Giessenbachstr.1, 85748 Garching, Germany (genzel@mpe.mpg.de, forster@mpe.mpg.de, linda@mpe.mpg.de, bender@mpe.mpg.de )*

*[2]Departments of Physics and Astronomy, University of California, 94720 Berkeley, USA*

*[3]Universitäts-Sternwarte Ludwig-Maximilians-Universität (USM), Scheinerstr. 1, München, D-81679, Germany (burkert@usm.uni-muenchen.de)*

*[4]Institute for Radio Astronomy in the Millimeter Range (IRAM), Rue de la Piscine, Grenoble, France*

*[5]Racah Institute of Physics, The Hebrew University of Jerusalem, Jerusalem 9190401, Israel*

*[6]Astronomy Department, Universidad de Concepción, Av. Esteban Iturra s/n Barrio Universitario, Casilla 160, Concepción, Chile*

*[7]Max-Planck Institute for Astrophysics, Karl Schwarzschildstrasse 1, D-85748 Garching, Germany*

*[8]School of Physics and Astronomy, Tel Aviv University, Tel Aviv 69978, Israel*

*[9]Osservatorio Astronomico di Padova, Vicolo dell'Osservatorio 5, Padova, I-35122, Italy*

*[10]Center for Computational Astrophysics, Flatiron Institute. 162 5th Avenue, New York, NY, 10010, USA*

*[11]Research School of Astronomy and Astrophysics, Australian National University, Canberra, ACT 2611, Australia*

*[12]ARC Centre of Excellence for All Sky Astrophysics in 3 Dimensions (ASTRO 3D)*

*[13]Department of Physics, University of Bath, Claverton Down, Bath, BA2 7AY, United Kingdom*

# ABSTRACT

We report high quality, Hα or CO rotation curves (RCs) to several $R_e$ for 41 large, massive, star-forming disk galaxies (SFGs), across the peak of cosmic galaxy evolution (z~0.67-2.45), taken with the ESO-VLT, the LBT and IRAM-NOEMA. Most RC41 SFGs have reflection symmetric RCs plausibly described by equilibrium dynamics. We fit the major axis position-velocity cuts with beam-convolved, forward modeling with a bulge, a turbulent rotating disk, and a dark matter (DM) halo. We include priors for stellar and molecular gas masses, optical light effective radii and inclinations, and DM masses from abundance matching scaling relations. Two-thirds or more of the $z \geq 1.2$ SFGs are baryon dominated within a few $R_e$ of typically 5.5 kpc, and have DM fractions less than maximal disks (median $\langle f_{DM}(R_e) \rangle = 0.12$). At lower redshift ($z<1.2$) that fraction is less than one-third. DM fractions correlate inversely with the baryonic angular momentum parameter, baryonic surface density and bulge mass. Inferred low DM fractions cannot apply to the entire disk & halo but more plausibly reflect a flattened, or cored, inner DM density distribution. The typical central 'DM deficit' in these cores relative to NFW distributions is ~30 % of the bulge mass. The observations are consistent with rapid radial transport of baryons in the first generation massive gas rich halos forming globally gravitationally unstable disks, and leading to efficient build-up of massive bulges and central black holes. A combination of heating due to dynamical friction and AGN feedback may drive DM out of the initial cusps.

# 1. INTRODUCTION

## *1.1 Rotation Curves as Probes of the Matter Distribution in the Local Universe*

Stellar and gas (HI, CO, Hα) rotation curves (RCs) have been a valuable tool for investigating the structure and mass distribution of galaxies, ever since the first stellar kinematic studies of the Milky Way in the early part of the last century (Kapteyn 1922, Oort 1927, c.f. Bland-Hawthorn & Gerhard 2016). This continued in the 1960s to 2000s in disk galaxies in the nearby Universe (Freeman 1970, Rubin & Ford 1970, Rogstad & Shostak 1972, Roberts & Rots 1973, Roberts & Whitehurst 1975, Rubin, Ford & Thonnard 1978, van der Kruit & Allen 1978, Sofue & Rubin 2001 Persic & Salucci 1988, Begeman, Broeils & Sanders 1991, Catinella et al. 2006, de Blok et al. 2008, Lelli, McGaugh & Schombert 2016). RCs with flat or positive slopes (circular velocity $v_c$ increasing with radius $R$) on scales of tens of kpc have since been one of the fundamental pillars of the dark matter paradigm, assuming that Newtonian physics applies on large scales (Ostriker, Peebles & Yahil 1974, Einasto, Kaasik & Saar 1974, Kent 1986, Courteau et al. 2014).

The *overall* RC (baryons plus dark matter) is determined by the mass fraction of the baryons in disk and bulge, to dark matter in the halo, $m_b=(M_{bulge}+M_{disk,*}+M_{disk,gas})/M_{DM}$, by the concentration of the halo, and by the specific angular momentum of the halo and the baryons. The angular momentum parameter of the baryons at the disk scale (Appendix A.2), $\lambda_{baryon}$, is often assumed to be similar to that of the dark matter at the virial scale ($\lambda_{DM}$). This simple assumption is supported by observations (Fall & Romanowsky 2013, Romanowsky & Fall 2012, Burkert et al. 2016), but theoretically not necessarily expected (Übler et al. 2014, Danovich et al. 2015, Teklu et al. 2015, Jiang et al. 2019). If the baryons are in a thin disk with an exponential distribution (Sersic index $n_S$=1) the half-mass (or effective) radius of the exponential disk is $R_e \sim (1.68/\sqrt{2}) \times \lambda_{baryon} \times R_{virial}$ (Mo, Mao & White 1998).

For the Milky Way (MW), the observed RC rises from the center to a peak at 6 kpc, then exhibits a shallow decline to 25 kpc, and then is a flat from 25-60 kpc (left panel of Figure A1, Battacharjee, Chaudhury & Kundu 2014, Bland-Hawthorn & Gerhard 2016, Eilers et al. 2019, Reid et al. 2019, Sofue 2020). The MW ratio of dark to total mass at $R_e$~4.5 kpc is about $f_{DM}(R_e)$= 0.38±0.1 (Bovy & Rix 2013, Bland-Hawthorn & Gerhard 2016) such that the MW disk is baryon dominated at $R_e$ and becomes dark matter dominated at the solar circle. The uncertainties of the various estimates are large enough not to rule out that the MW is a 'maximal' disk ($f_{DM}(R_e) \equiv$ 0.28, van Albada et al. 1985). The MW RC is typical of other massive, bulged disks at z~0 (Sofue & Rubin 2001, van der Kruit & Freeman 2011). Very massive ($v_c$~250-310 km/s) disks with large bulge to total ratios have $f_{DM}(R_e)$=0.25±0.15 but are rare (Barnabe et al. 2012, Dutton et al. 2013). Otherwise typical disk galaxies in the local Universe are more dark matter dominated with $f_{DM}(R_e)$~0.5-0.9. The most dark matter dominated disks tend to have low baryonic mass and circular velocity (Martinsson et al. 2013a,b, Courteau & Dutton 2015).

### *1.2 Rotation Curves in high-z Star Forming Disks*

The cosmic star formation density peaked 5-11 Gyr ago (z~1-2.5) (Madau & Dickinson 2014), during which galaxy halos containing Milky-Way mass galaxies first formed in large numbers (e.g. Mo & White 2002). Over the past two decades high-throughput, adaptive optics assisted, near-integral field spectrometers (IFS), such as SINFONI on the ESO-VLT (Eisenhauer et al. 2003, Bonnet et al. 2004), or OSIRIS on the Keck telescope (Larkin et al. 2003), and seeing limited, multiplexed IFSs, such as KMOS at the VLT (Sharples et al. 2012), have become available on 8-10 m telescopes. With these IFSs it is now possible to carry out deep, velocity resolved (FWHM~80-120 km/s) spectroscopic imaging of Hα in z~0.6-2.6 disk galaxies on the star forming 'main sequence (MS)' (Whitaker et al. 2012, 2014, Speagle et al. 2014).

Over the past decade we have carried out two main IFS surveys of high-z galaxy kinematics, SINS & zC-SINF with SINFONI (Genzel et al. 2006, Förster Schreiber et al. 2006, 2009, 2018), and KMOS$^{3D}$ with KMOS (Wisnioski et al. 2015, 2019). In total we have assembled ~850 IFS data sets of MS star forming galaxies (SFGs) across the cosmic star/galaxy formation peak, covering the mass range from log($M_*$/M$_\odot$)=9.5-11.5. Between 60-80% of the MS galaxies at these masses are rotation dominated, turbulent and thick disks with $v_{rot}(R_e)/\sigma_0$~3-10 (Wisnioski et al. 2015, 2019, Simons et al. 2017). Here $v_{rot}(R_e)$ is the inclination and beam smearing corrected intrinsic rotation velocity of the disk at $R_e$, and $\sigma_0$ is the average velocity dispersion of the (outer parts of the) disk, after removal of beam-smeared rotation, instrumental line broadening, and other orbital motions. In our SINS and KMOS$^{3D}$ surveys, we have emphasized deep integrations, for high quality data on individual galaxies, rather than optimizing the number of observed galaxies – an approach that has enabled excellent quality high-z RCs.

In addition RCs of high-z galaxies can now also be obtained from millimeter and submillimeter emission lines, such as CO and [CII] (e.g. Genzel et al. 2013, Übler et al. 2018). With the availability of such high quality high-z RCs, it is possible to go beyond the mere demonstration of rotationally dominated motions (Förster Schreiber & Wuyts 2020), to exploiting RCs for studying mass distributions in more detail.

### *1.3 Summary of Published High-z Rotation Curve Studies*

Wuyts et al. (2016) modelled the *inner disk kinematics* with the first and second velocity moments of Hα along the kinematic major axis in 240 z=0.6-2.6 SFGs from the KMOS$^{3D}$ survey (Wisnioski et al. 2015, 2019). They compared the dynamical masses within $R_e$ with the sum of stellar and gas masses inferred from multi-band optical/IR photometry, HST imaging, and molecular gas scaling relations (e.g. Tacconi et al 2018). Wuyts et al. (2016) found average baryon fractions of $f_{baryon}(<R_e)$~60%, increasing with redshift to ~90% at z~2.2. Übler et al. (2017) inferred a similar global increase in baryon fractions from z<1 to z>2 for 135 'best disks' in a '*Tully-Fisher offset analysis (TFA)*' of KMOS$^{3D}$

SFGs. Turner et al. (2017) carried out a TFA study in 1200 z=0-3 SFGs in KMOS-KDS and literature samples, and Tiley et al. (2016, 2019a) in 210 z=1 SFGs in KMOS-KROSS. All three TFA studies find similar offsets as a function of z. Van Dokkum et al. (2015) observed integrated Hα linewidths in 25 compact SFGs with MOSFIRE at the Keck telescope and concluded that these widths are consistent with a Keplerian fall-off of circular velocities, indicative of masses dominated by dense stellar components.

Genzel et al. (2017) presented deep SINFONI and KMOS$^{3D}$ Hα imaging spectroscopy for 6 very massive, bulgy z=0.9-2.4 SFGs ($\log(M_{baryon}/M_\odot)$=11.1-11.3), tracing individual RCs to 1.5-3 Re for the first time at these redshifts. In four or five of the six SFGs the RCs decline beyond a maximum near $R_e$. Genzel et al. (2017) interpreted these radial drops as being due to a combination of asymmetric drift at high velocity dispersions and large baryon fractions within $R_e$ ($f_{DM}(R_e)$=0→30%).

To test whether the results of Genzel et al. (2017) are representative of the overall population of high-z massive SFGs, Lang et al. (2017) stacked 101 massive, large, rotationally supported SINFONI and KMOS$^{3D}$ SFGs with median z>=1.52 and $\log(M_*/M_\odot)$=10.6, which had robustly detected turnover radii. They normalized individual RCs by turnover radius and corresponding velocity before stacking. The resulting average RC shows a radially decreasing RC, in agreement with Genzel et al. (2017). Lang et al. (2017) also showed that a second method of 'stellar light normalization' using the effective radius $R_e$ and Sérsic index *n* from *HST* NIR imaging to estimate the turnover radius gave comparable results. Using four different approaches to estimate dark matter fractions near $R_e$, the results of Genzel et al. (2017), Lang et al. (2017), Übler et al. (2017) and Wuyts et al. (2016) all indicate that *massive rotating disks have low dark matter fractions within several $R_e$ at z>1-2.5. Dark matter fractions increase towards lower redshift*, in agreement with the local Universe results described above.

***Other Studies.*** Price et al. (2016, 2020) analyzed Hα slit kinematics from the Keck MOSDEF survey for 681 z~1.4-3.8 SFGs, for another independent look at the inner disk kinematics. Their results are in excellent agreement with Wuyts et al. (2016). Tiley et al. (2019b) compiled 1500 z=0.6-2.2 SFGs from the KROSS (551 galaxies with median z=0.85 and $\log(M_*/M_\odot)$=10.0), KGES (228 galaxies with median z=1.5 and $\log(M_*/M_\odot)$=10.3), KMOS$^{3D}$ (145 galaxies with median z=2.3 and $\log(M_*/M_\odot)$=10.3) surveys, as well as MUSE observations (96 galaxies with median z=0.67 and $\log(M_*/M_\odot)$=9.8). Tiley et al. confirmed dropping RCs similar to Lang et al. (2017) for all their sub-stacks when adopting a similar 'self-normalization' methodology based on the RC turnover, but found instead flat or even rising average RCs when normalizing by the observed velocity at three times the disk scale length, $3R_d$, assuming exponential light distributions (and accounting for spatial beam smearing). This illustrates the impact of different approaches, and possibly sample differences, in stacking and highlights the need to study individual RCs.

# 2. A Second Generation Data Set of High Quality RCs: RC41

### *2.1 Summary of the RC41 Sample*

In this paper we present 41 high quality, individual z=0.65-2.45 RCs (35 new RCs and the 6 RCs from Genzel et al. 2017) that we have collected and analyzed since Genzel et al. (2017), which we will refer to as the 'RC41' sample. In following up on Genzel et al. (2017) our main goal was to increase statistics, and to expand the sample to lower masses and lower redshifts. With RC41, we also better sample the overall MS population across the peak of cosmic galaxy formation than in the earlier work. Of the 41 RCs, 17 come from SINFONI at the ESO-VLT in both non-AO and AO (14) modes. For seven we combined SINFONI observations with data obtained with the KMOS multi-IFS instrument at the VLT. For eight we used KMOS data only. We also compare and combine Hα- and molecular gas (CO-based) RCs in 2 galaxies. Genzel et al. (2013) and Übler et al. (2018) have shown excellent agreement of CO (from IRAM-NOEMA) and Hα (from LBT-LUCI) RCs for two SFGs with comparable spatial resolutions. In RC41, we now add 7 new NOEMA-CO RCs. RC41 on-source integration times varied from 4 to 56 hours, with a median of 16 hours, and a total of 800 on-source hours, equivalent to at least 1000 total telescope hours on the ESO-VLT, on the LBT and on NOEMA. Figure 1 shows the locations of the RC41 sources in the planes of log $M_*$ vs. MS-offset ($\delta MS$=log($SFR/SFR(MS,z)$)), effective radius $R_e$ of optical (5000 Å) stellar continuum, baryonic (sum of stellar and gas) mass surface density within $R_e$, log $\Sigma_{baryon}$, and bulge mass $M_{bulge}$. Throughout this paper filled blue and red circles denote RC41 galaxies in the redshift slices 0.65-1.2, and 1.2-2.45. Table 1 summarizes the relevant basic 'input' properties of the RC41 sample.

Our VLT targets are primarily selected from and benchmarked to the 3D-HST and CANDELS surveys (Skelton et al. 2014; Momcheva et al. 2016; Grogin et al. 2011; Koekemoer et al. 2011). Additional targets come from the Steidel et al. (2004) 'BX-BM' Lyman-break surveys, the GMASS Spitzer survey (Cimatti et al. 2008), the zCOSMOS-Deep survey of the COSMOS field (Lilly et al. 2007, Scoville et al. 2007), and BzK galaxies from the Deep3a field (Kong et al. 2006). The observations were taken as part of the SINS/zC-SINF Hα survey of 92 galaxies with SINFONI (Förster Schreiber et al. 2009, 2018; Mancini et al. 2011) and the KMOS$^{3D}$ survey of 740 galaxies with KMOS (Wisnioski et al. 2015, 2019), each with an Hα detection fraction of about 80% (and very small overlap of 18 sources observed in both surveys). At NOEMA in the northern hemisphere, the targets were observed in the PHIBSS 1 & 2 CO surveys (Tacconi et al. 2010, 2013, 2018; Übler et al. 2018; Freundlich et al. 2019), selected and benchmarked from the All-Wavelength Extended Groth Strip International Survey (abbreviated here as EGS: Davis et al. 2007; Noeske et al. 2007), with spectroscopic redshifts from DEEP2 (Newman et al. 2012). Additional PHIBSS targets were taken from the zCOSMOS-Deep survey (Lilly et al. 2007), and the rest from 3D-HST in the GOODS-N, EGS and COSMOS fields. Most of the 180 PHIBSS targets were observed in compact array configurations as the focus was on source-integrated properties, with an 82% detection fraction; 12 of the CO-detected galaxies were followed up in extended configurations to more fully resolve their line emission.

***Sample Selection.*** From the full Hα and CO parent sample of 800 unique detected galaxies, with one exception, we selected rotationally supported ($v_{rot}/\sigma_0$>2.3) SFGs with stellar masses 9.5≤log($M_*/M_\odot$)≤11.5, redshifts 0.65 ≤ z ≤ 2.5, and near the MS, -0.6 ≤ $\delta MS$ ≤1. Given the atmospheric transparency in the near-infrared and millimeter ranges, this splits the data into three redshift slices, 0.65 ≤ z ≤ 1.2, and 1.2 ≤ z ≤ 1.6 and 2.02 ≤ z ≤ 2.5. These selections assure a fairly homogeneous coverage of the star-forming disk population around the MS in the stellar mass - star formation rate plane in each redshift slice. The kinematic criterion removes about one-fourth of the detected galaxies; of the remaining objects, 86% (514) satisfy the stellar mass, redshift, and MS offset cuts.

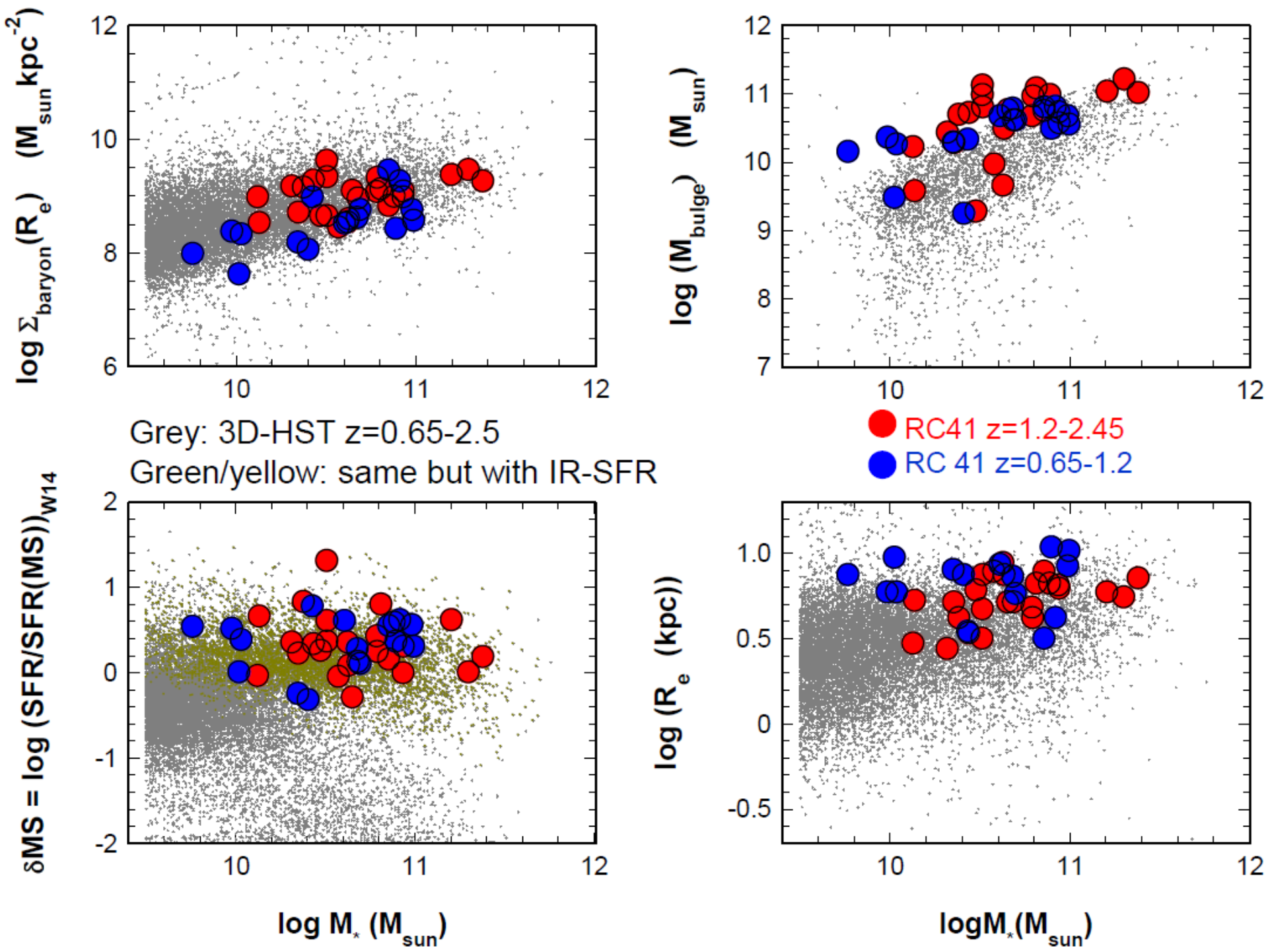

Figure 1. Locations of the RC41 SFGs in the planes stellar mass vs. MS-offset (bottom left), stellar mass vs. effective radius (5000 Å) (bottom right), stellar mass vs. baryonic surface density (stars plus gas) (upper left) and stellar mass vs. bulge mass (upper right). Filled circles denote the location of the 41 galaxies (in redshift bins 0.65-1.2 (blue), 1.2-2.45 (red)) relative to the galaxies in the 3D-HST parent catalog (Skelton et al. 2014; Momcheva et al. 2016), a near-IR grism survey with the *Hubble Space Telescope* (*HST*) in the CANDELS *HST* imaging survey fields (Grogin et al. 2011; Koekemoer et al. 2011) and with extensive X-ray to radio multi-wavelength data. In the bottom left panel grey crosses denote galaxies for which only SED-based SFRs are available, while green/yellow crosses denote galaxies where mid- or far-infrared based star formation rates are available in the Wuyts et al. (2011a) 'ladder' scheme of star formation indicators.

We retained only spatially well resolved data sets, necessary to derive RCs. This restriction culls compact SFGs ($R_e$< 2 kpc), given the FWHM angular resolution of our data sets, which range from 0.25" for AO data sets to 0.8" for the lowest resolution KMOS and NOEMA data, leaving 415 objects. In addition, we required significant line detection at $R \geq R_e$ and further eliminated sources for which residual OH sky line emission affected the Hα line profile over part of the galaxy, leading to the final RC41 sample that represents the best cases in terms of S/N, radial coverage, and quality among our data sets. The statistics are best at the massive tail of the star-forming MS population (bottom left of Figure 1) and the median optical effective radius in RC41 is 5.7 kpc. The one exception noted above is a more actively star-forming disk at $\delta MS$ ~ 1.25, which otherwise meets all other criteria applied including the particularly stringent ones of well-detected and well-resolved line emission out to large radii. Taken together, the two size and line emission extent criteria result in the largest cut and *bias the RC41 sample to larger disks* (bottom right panel of Figure 1), which is an inevitable but acceptable consequence of wanting to push high-z RCs to ≥0.1 $R_{virial}$. This means that the baryonic surface densities of the disks in RC41 are at the *low density tail* of the overall $\Sigma_{baryon}$ distribution (upper left panel of Figure 1). With the main conclusions of this work in mind, it is fair to say that this selection is unlikely to bias dark matter fractions downward. As discussed in van Dokkum et al. (2015) and Wuyts et al. (2016), more compact SFGs, missing from RC41, have higher baryonic surface densities and are likely to be more baryon dominated than RC41 galaxies.

***Dynamical Modeling***. As in our earlier work (Genzel et al. 2006, Wuyts et al. 2016, Burkert et al. 2016, Genzel et al. 2017, Lang et al. 2017, Übler et al. 2017, 2018) we use forward modeling from a parameterized, input mass distribution to establish the best fit models for a given Hα or CO data set. This mass model is the sum of an unresolved passive bulge (not emitting in Hα or CO), a rotating flat disk of Sersic index $n_S$, effective (half-light) radius $R_e$ and (constant) isotropic velocity dispersion $\sigma_0$, and a surrounding halo of dark matter. As discussed in more detail in the above cited references and in Appendix A, we compute from this mass three dimensional data cubes ($I(x,y,v_z)$) of the disk gas, convolved with a three dimensional kernel describing the instrumental point spread function PSF ($\delta x, \delta y, \delta v_z$) of our measurements. We then compare directly this 'beam smeared' model to the observed data , and vary model parameters to obtain best fits. The most common approach (taken in this paper) is to extract velocity centroids (from Gaussian fits) and velocity dispersion cuts from a suitable software slit along the dynamical major axis of the galaxy, for both the model and measurement cubes. We either use a constant software slit width (typically ~1-1.5 FWHM of the data set), or for low inclination targets where the iso-velocity contours fan out, a fanned slit width (5-10 degree opening angle) (e.g. van der Kruit & Allen 1978). This 1D method is applicable to *all* RC41 galaxies, since most of the RC information is contained along this major axis (c.f. Genzel et al. 2006, Genzel et al. 2017). For the highest resolution data sets, or very large galaxies, we obtain additional information by investigating the model and measurement 2D velocity and velocity dispersion maps, or even comparing the individual spaxels in the data cubes for a full 3D analysis. We discuss these 2D and 3D analyses for suitable RC41 galaxies in Paper 2 of this series (S. Price, T. Shimizu, et al., in preparation), and refer the reader for the full detailed discussion to that paper.

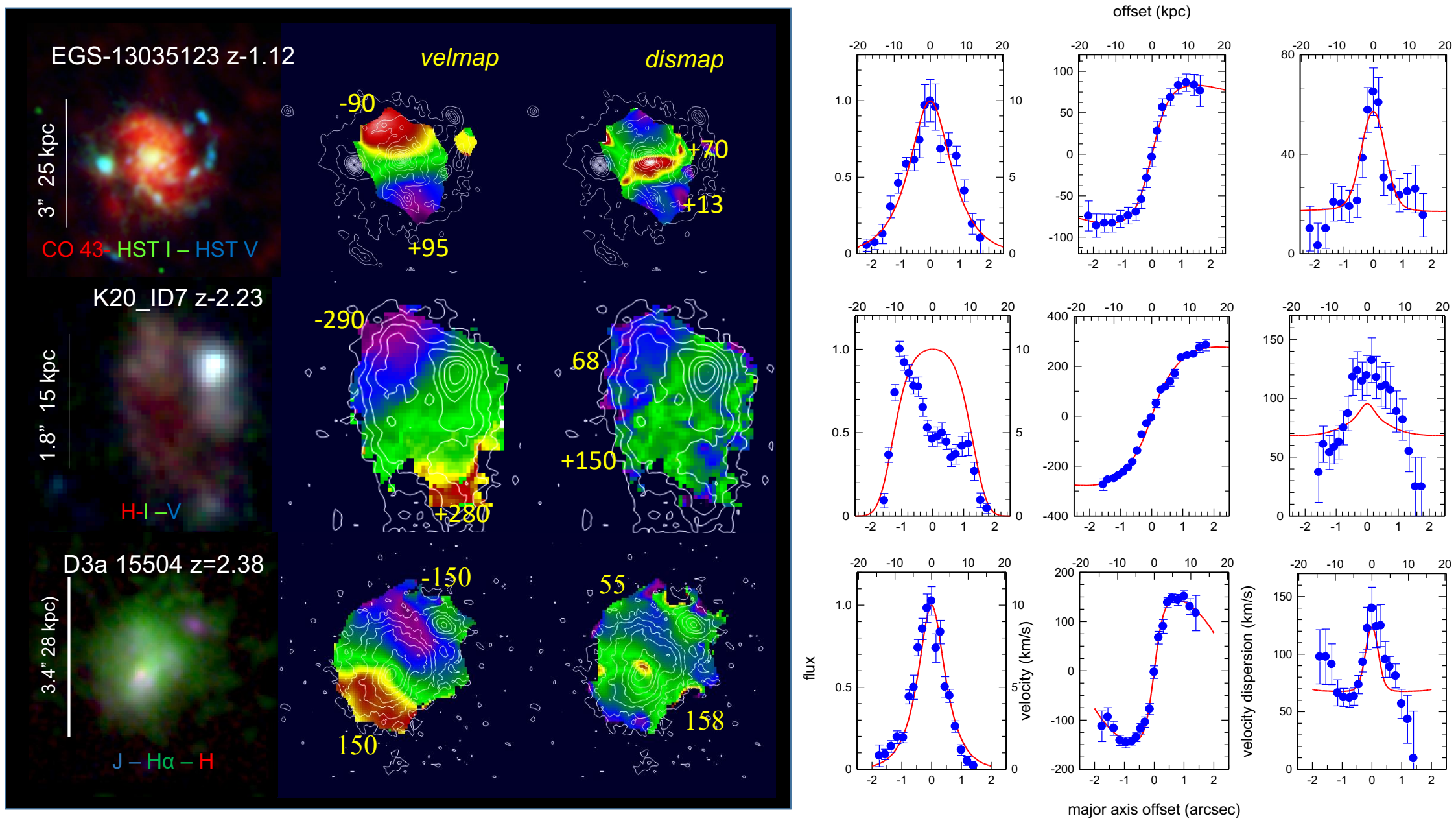

Figure 2. Three examples of RC41 SFGs with different kinematic properties. From left to right: two to three-band, HST or Hα images, Hα or CO velocity and velocity dispersion distributions (on top of HST H- or I-band map in contour form), and the 1D-major axis cuts of Hα or CO intensity, velocity and velocity dispersion (blue filled circles with 1σ uncertainties). The red curves are the best-fit models inferred from these data. The three galaxies show different RC shapes. Beyond $R_e$ the top galaxy has a flat rotation curve, the central galaxy has a rising rotation curve, and the bottom galaxy a declining rotation curve. These different RC shapes give important constraints on the relative amounts or baryonic and dark matter in the outer disks.

Figure 2 shows three examples of the full data sets for each galaxy, along with the best fit model (red line in the right three panels), obtained from the analysis described in Appendix A. Figure 3 shows the inclination corrected model RCs of the baryons, dark matter and total mass, for the three galaxies in Figure 2. For each, we have two or three band HST imagery, plus high quality Hα or CO brightness distributions, and first and second moment kinematic maps. The bottom (Figure 2) case (left panel in Figure 3) is a strongly baryon dominated z~2 galaxy (also contained in Genzel et al. 2006, 2017), where the overall intrinsic circular velocity curve drops with radius, and where turbulent pressure corrections (asymmetric drift, A1) are very important and lead to the precipitous drop in rotation velocity. In contrast, the galaxy shown in the middle panels is a dark matter dominated z~2 system, with a rising rotation curve. The top panels in Figure 2 (right panel in Figure 3) feature a larger, z~1 galaxy with a flat rotation curve.

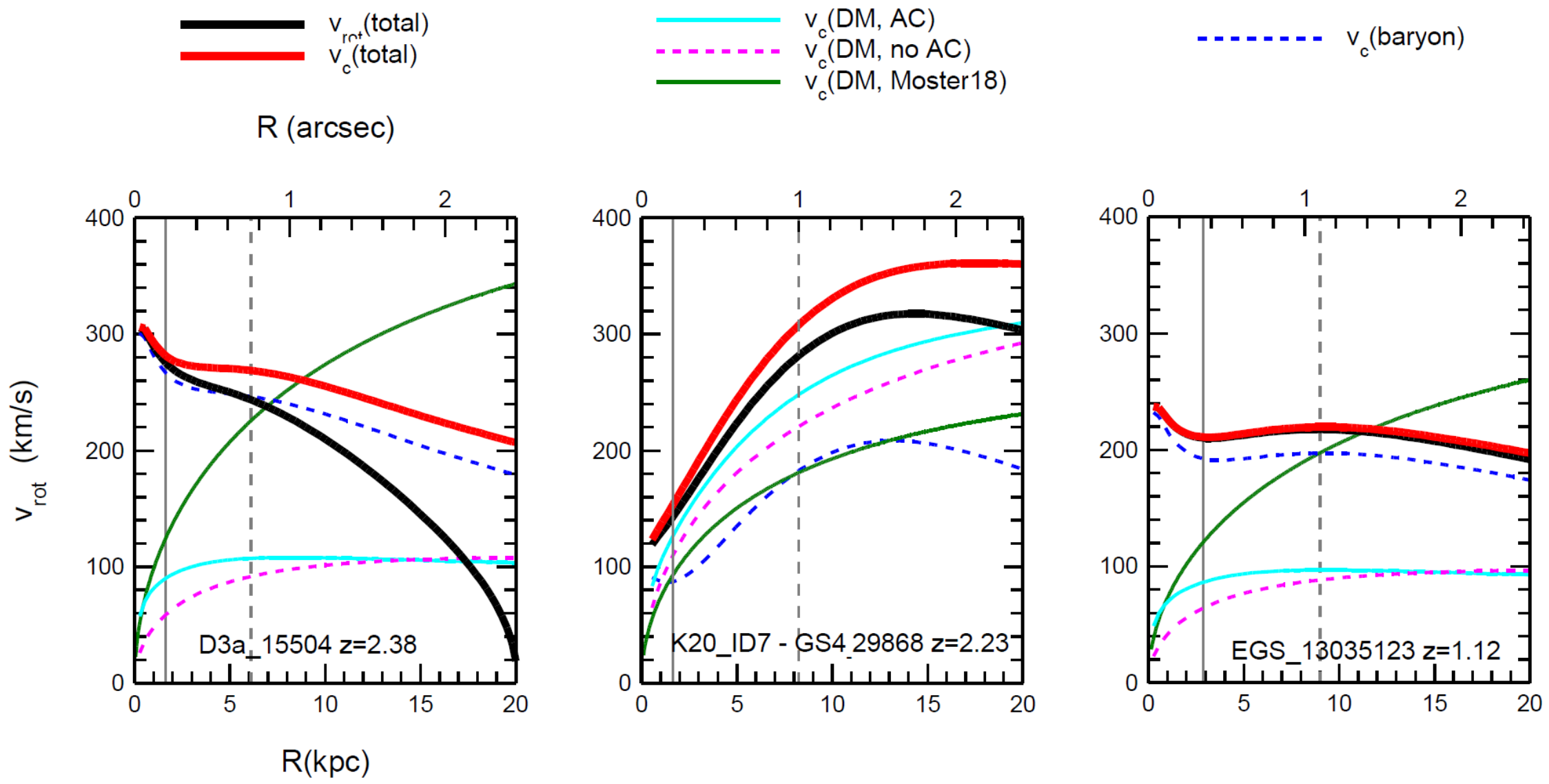

Figure 3. Inclination corrected, intrinsic RCs for baryons, dark matter, and total, from the dynamical fits for the three galaxies in Figure 2. The red lines denote the total circular velocity $v_c(R)$, the black curves give $v_{rot}$ (R), correcting $v_c(R)$ for the effect of turbulent pressure (asymmetric drift), the dashed blue curves denote $v_{baryon}$ *(R)* (disk and bulge combined), the dashed pink and continuous cyan curves are the fitted dark matter RCs $v_{DM}(R)$, without and with adiabatic contraction (Mo, Mao & White (1998)), respectively. Finally the continuous green curve is $v_{DM}(Moster)$, obtained from an NFW model without adiabatic contraction and a total mass equal to $M_{DM}(Moster)$ (Moster et al. 2018). The vertical dashed grey line marks $R_e$, and the vertical solid grey line the HWHM of the measurements.

The forty-one best major axis position-velocity diagrams make up RC41. In this paper, we focus on the analysis of the 1-D cuts of the Hα/CO data on the kinematic and structural major axis. Figure 4 summarizes all of the observed RCs of the RC41 sample (black filled circles with rms uncertainties), along with the projected (and beam-convolved) best-fit, model RCs (red curves). Table D1 in Appendix D summarizes the inferred properties of the RC41 galaxies obtained from the analysis. Paper 2 will examine the 2-D distributions, and analyze non-axisymmetric distributions and/or radial motions.

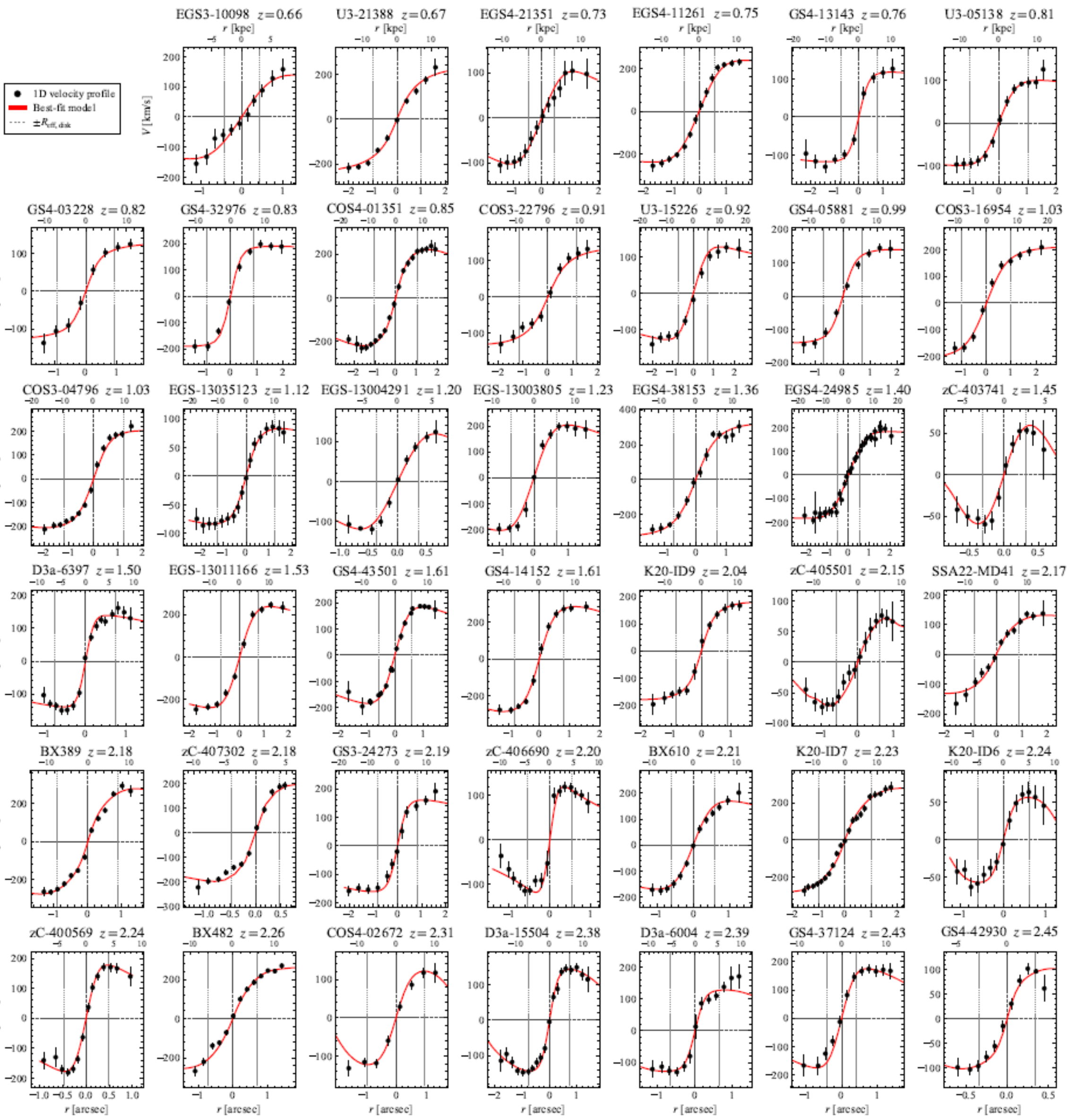

Figure 4. Line velocity centroids (from Gaussian fits) as a function of position offset ('position-velocity cuts') along the galaxy's major axis (0 is the dynamical center) in arcsec (below) and in kpc (above), for the 41 RC41 sample galaxies, ordered in ascending order of redshift, from top left to bottom right. Black points (and ±1σ errors) are the measurements, and the red curves are the best-fit, beam-convolved and projected model fits (same as in Figure 3). The name and redshift of each galaxy is listed above the p-v diagram. The two grey vertical lines symmetrically on either side of the dynamical center denote the effective radius. We find a variety of RC shapes: peaked or falling, flat and increasing up the outermost radius. Very few of the 41 RCs show significant reflection asymmetries indicative of perturbations and warps.

# *3.* Results

## *3.1 Equilibrium vs. Perturbations*

***Shapes and Symmetry of the Observed Rotation Curves.*** Figure 4 shows that the RCs of RC41 have a variety of shapes. We make a qualitative assessment of their shapes: About half (17 of 41) are falling, or exhibit a clearly defined maximum and begin to drop beyond that. About half of these again are at z>2.1. Eight of the 17 exhibit substantial drops in rotation velocity (factors of 1.2-2 from the peak to the outermost points at ~2-3 $R_e$). One-third (15 of 41) of the RC41 SFGs have flat RCs. RCs rising to the outermost measurement point (typically 2 $R_e$ and in three cases 2.5 $R_e$) are relatively rare (9 of 41, or 20%) but are seen across the full redshift range. For details of the modelling of the individual RCs, including the velocity dispersion, Hα flux cuts and HST imagery, we refer the reader to Appendix A.

We determined how many of the 41 RCs show significant deviations from reflection symmetry around the dynamical center[1], and whether the rotation curves are flat, rising or reach a peak and turn down, or even fall substantially (by a factor of >1.1) from the maximum to the outermost point. Deviations from point symmetry could be indications of significant perturbations by nearby satellite galaxies, or of warps in the outer galaxy disk. We find that for 33 of the 41 RCs (82.5%) a parabola fit determined from the RC on one side of the galaxy also fits the other side with $\chi^2 \leq 1.2$ (Figure 5). Only 3 (5) SFGs have $\chi^2 > 2$ ( ≥ 1.5). The reflection symmetry of our RC41 sample is quite striking. Simulated high-z galaxies in the currently highest resolution cosmological hydro-simulations, Illustris TNG50, are far less symmetric (Pillepich et al. 2019, Übler et al. 2020).

[1] note that the zero points in velocity and position-offset along the major axis are free 'nuisance' parameters, which need to be established from the data. For the former, we use the average of peak velocities on either side of the center as the 'systemic' velocity. The combination of the position offsets at the systemic velocity, at the maximum of velocity dispersion and, if available, the position of a central bulge then define the latter.

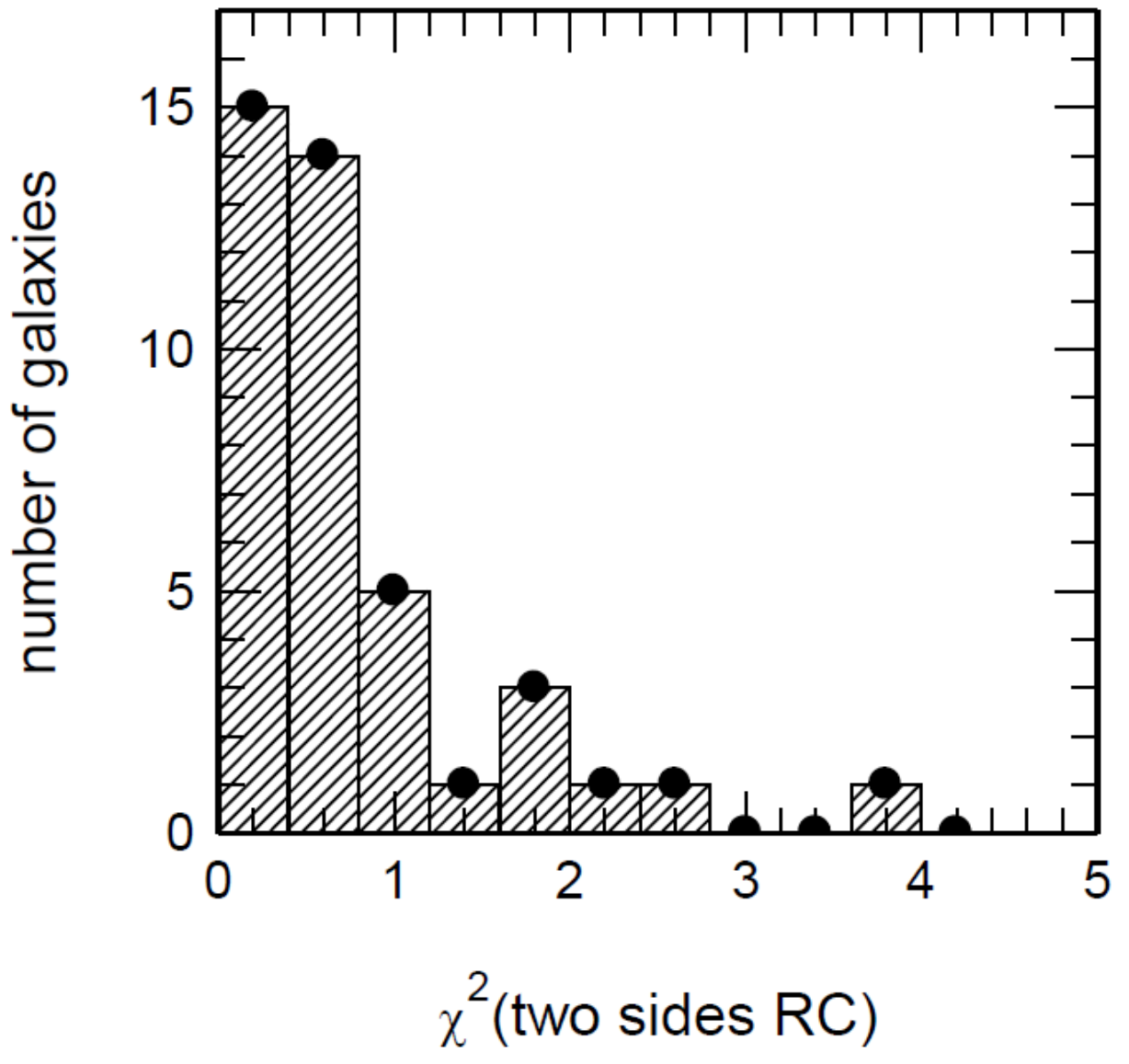

Figure 5. Reflection symmetry of the two sides of each rotation curve for all 41 RC41 galaxies. For each galaxy we fold the RC at the galaxy center (bulge, or point of maximum velocity dispersion or largest velocity gradient), and fit a parabola ($|v| = a + b*(p - p_0)^2$) for the RC on one side, and then evaluate the $\chi^2$ of the fit of this equation to the data (and their uncertainties) on the other side. Typically, 3-5 data points are available for this fit. Only 3 galaxies exhibit significant deviations from mirror symmetry of the RC ($\chi^2>2$), 34 have $\chi^2\leq1$.

***Environment and Perturbations***. Previous observations and simulations indicate that the rate of galaxy interactions and mergers increases rapidly with redshift (LeFevre et al. 2000, Conselice et al. 2003, Wetzel et al. 2009, Lopez-Sanjuan et al. 2010, Rodriguez-Gomez et al. 2015, Mantha et al. 2018, Cibinel et al. 2019, Pillepich et al. 2019). For instance, the most recent TNG50 simulations find that at z>2 even moderately massive galaxies are asymmetric with disturbed disk rotation (Pillepich et al. 2019, Übler et al. 2020). What can we say about the environment and the level of perturbations for the RC41 sample?

Table 1 summarizes our findings. In RC41 there are only two major mergers/interacting systems (mass ratio <3:1), one early stage (GS4_13143, z=0.76, separation 2", or 15 kpc), and one late-stage, with a prominent tidal tail surrounding a compact, but apparently well settled, compact bulge-disk, EGS_13004291 (z=1.197, Tacconi et al. 2013). There are 8 systems (7 at z>2.1) where the Hα velocities show evidence for a clear dynamical interaction with a smaller mass galaxy – in those cases the galaxy mass ratios range from 6:1 to 30:1. Most of these are probably satellites or minor mergers; there are three galaxy groups with 2 or 3 members. There are 5 additional cases of a central galaxy with a nearby projected companion (mass ratio 5:1 to 50:1) but no spectroscopic evidence for physical association. In summary, 10-15 out of the 41 systems show evidence for companions or satellites. For comparison, an estimated 5-10% of all KMOS$^{3D}$ galaxies (z=0.6-2.6; Förster Schreiber, priv. comm.) have interacting satellites/minor mergers with mass ratios <30:1. This is consistent with RC41 since the RC41 data are deeper, thus making physical interactions easier to detect. RC41 also has relatively more z>2 galaxies than the full KMOS$^{3D}$ sample.

Such interactions and their resulting tidal forces can perturb the motions in the central galaxy, and/or induce warps. The last two columns in Table 1 provide approximate estimates of the impact of the satellites/minor mergers on the gas motions at $\geq R_e$ in the central galaxies. In Column 15 we compute the relative change in velocity of a particle in the central galaxy A, normalized to the circular velocity at $R_e$, caused by the gravitational pull from the satellite B during the 'encounter time' $\delta t \sim 1\text{-}1.5 \times R_{AB}/\Delta v$. Here $R_{AB}$ is the projected separation between A and B and $\Delta v$ is the observed encounter velocity from column 14. The estimate in column 15 is a lower limit for the effect of B on A during a one-time encounter on an unbound or barely bound interaction. In column 16 we compute the ratio of $R_e(A)$ to the 'tidal', or Jacobi (or Hills) radius of a particle in galaxy A, that is strongly perturbed by B, given the separation $R_{AB}$. If this ratio exceeds unity, then a particle in the central galaxy at $R_e$ will be strongly perturbed in its long-term orbit due to tidal forces from orbiting B (Binney & Tremaine 2008, chapter 8, eq. 8.14). This estimate gives an upper limit to the effect of tidal forces of the minor merger or satellite, as it assumes a stable, near circular orbit. Inspection of columns 15 and 16 show that given the mass ratios and separations of the interaction partners, the tidal perturbations in RC41 are modest (~0.1) and could on average explain ~20 km/s perturbations in the rotation curves. This can be significant in some galaxies but is comparable to the measurement uncertainties in the outer disks of most of our galaxies. In further support of this conclusion, of the 10-11 galaxies with spectroscopic evidence for interactions, only two have somewhat asymmetric rotation curves, with $\chi^2 > 2$ (Figure 5). We cautiously conclude that for most of the RC41 sample there is no convincing evidence that perturbations and environmental interactions affect the kinematics.

| | 1 | 2 | 3 | 4 | 5 | 6 | 7 | 8 | 9 | 10 | 11 | 12 | 13 | 14 | 15 | 16 |
|---|---|---|---|---|---|---|---|---|---|---|---|---|---|---|---|---|
| | galaxy | z | morphology | inclination | v_c (R_e) km/s | R_e kpc | kpc/ arcsec | logM_* HST input (M_sun) | B/T optical light | B/T dynamical | satellite/ companion/ merger | distance (arcsec) | light/mass ratio central/satellite | Δv (km/s) satellite-central | δv/v_A in interaction | R_J/ R_AB |
| 1 | EGS3_10098/ EGS4_30084 | 0.66 | outer spiral/ring+bulge | 31 | 326 | 3.0 | 6.9 | 11.1 | | 0.60 | | | | | | |
| 2 | U3_21388 | 0.67 | N3079 disk+bulge | 82 | 230 | 7.0 | 7.0 | 10.8 | 0.39 | 0.05 | | | | | | |
| 3 | EGS4_21351 | 0.73 | bulge & ring | 47 | 187 | 3.3 | 7.2 | 10.9 | 0.21 | 0.46 | | | | | | |
| 4 | EGS4_11261 | 0.75 | bulge & clumpy disk | 60 | 327 | 4.0 | 7.3 | 11.3 | | 0.50 | | | | | | |
| 5 | GS4_13143 | 0.76 | spiral +clumpy nuclear disk | 74 | 156 | 5.6 | 7.4 | 9.8 | | 0.70 | interaction | 2" ENE | H: 2.2:1 | 165 | 0.11 | 0.15 |
| 6 | U3_05138 | 0.81 | bulge & disk | 55 | 145 | 7.5 | 7.5 | 10.2 | 0.03 | 0.50 | companion | 1.3"NNW | H: 50:1 | | | |
| 7 | GS4_03228 | 0.82 | bulge & disk | 78 | 146 | 7.1 | 7.6 | 9.5 | | 0.80 | | | | | | |
| 8 | GS4_32976 | 0.83 | bulge & disk | 68 | 249 | 6.8 | 7.6 | 10.4 | 0.09 | 0.90 | | | | | | |
| 9 | COS4_01351 | 0.85 | bulge+ring | 68 | 290 | 8.0 | 7.7 | 10.7 | 0.02 | 0.24 | | | | | | |
| 10 | COS3_22796 | 0.91 | small bulge & big ring | 58 | 145 | 9.0 | 7.8 | 10.3 | 0.10 | 0.15 | | | | | | |
| 11 | U3_15226 | 0.92 | bar-wide spiral | 50 | 194 | 5.5 | 7.9 | 11.1 | 0.24 | 0.55 | | | | | | |
| 12 | GS4_05881 | 0.99 | clumpy disk +extermely red bulge? | 60 | 200 | 5.6 | 8.0 | 9.8 | | 0.85 | | | | | | |
| 13 | COS3_16954 | 1.03 | bulge+spiral arms | 49.5 | 282 | 8.1 | 8.1 | 10.7 | 0.08 | 0.50 | companion? | 3" WNW | H: 26:1 | | | |
| 14 | COS3_04796 | 1.03 | spiral+bulge | 50 | 269 | 9.7 | 8.1 | 10.8 | 0.04 | 0.18 | satellite | 1.1" ENE | H 13:1 | 200 | 0.08 | 0.25 |
| 15 | EGS_13035123 | 1.12 | spiral+bulge | 24 | 220 | 10.2 | 8.2 | 11.2 | 0.06 | 0.20 | | | | | | |
| 16 | EGS_13004291 | 1.20 | clumpy disk,+ large bulge, strong tidal arm | 27 | 354 | 3.0 | 8.3 | 11.0 | | 0.61 | late stage merger | | | | | |
| 17 | EGS_13003805 | 1.23 | spiral+big extincted bulge | 37 | 393 | 5.6 | 8.3 | 11.2 | | 0.29 | | | | | | |
| 18 | G4-38153 | 1.36 | edge on ring | 75 | 356 | 5.9 | 8.4 | 10.4 | | 0.16 | | | | | | |
| 19 | G4_24985 | 1.40 | extincted bulge &disk | 40 | 300 | 4.6 | 8.4 | 10.9 | | 0.40 | | | | | | |
| 20 | zC_403741 | 1.45 | bulge & ring | 28 | 204 | 2.6 | 8.5 | 10.7 | | 0.68 | | | | | | |
| 21 | D3a_6397 | 1.50 | bulge & ring | 30 | 308 | 6.4 | 8.5 | 11.1 | | 0.57 | | | | | | |
| 22 | EGS_13011166 | 1.53 | small bulge & clumpy spiral arms | 60 | 348 | 6.3 | 8.6 | 11.1 | | 0.55 | | | | | | |
| 23 | GS4_43501 (GK 2438) | 1.61 | bulge & ring | 62 | 257 | 4.9 | 8.5 | 10.6 | 0.06 | 0.40 | | | | | | |
| 24 | GS4_14152 | 1.62 | bulge & disk | 55 | 397 | 6.8 | 8.5 | 11.3 | | 0.23 | | | | | | |
| 25 | K20_ID9-GS4_27404 | 2.04 | small bulge & ring | 48 | 250 | 7.1 | 8.3 | 10.6 | 0.14 | 0.30 | | | | | | |
| 26 | zC_405501 | 2.15 | bulge & edge on clumpy disk | 75 | 139 | 5.0 | 8.3 | 9.9 | | 0.07 | interacting group with 3 members | N +1.7" SSE -1.75" | H: 2.7:1 5.4:1 | | | |
| 27 | SSA22_MD41 | 2.17 | ring | 72 | 189 | 7.1 | 8.3 | 9.9 | | 0.05 | | | | | | |
| 28 | BX389 | 2.18 | edge on ring | 76 | 331 | 7.4 | 8.3 | 10.6 | | 0.30 | satellite | 0.7" SSW | m: 13:1 | 180 | 0.12 | 0.27 |
| 29 | zC_407302 | 2.18 | disk | 60 | 280 | 4.0 | 8.3 | 10.4 | | 0.50 | satellite | 0.55" N | m: 10:1 | 90 | 0.19 | 0.22 |
| 30 | GS3_24273 | 2.19 | extincted bulge & clumpy disk | 60 | 222 | 7.0 | 8.3 | 11.0 | 0.15 | 0.80 | | | | | | |
| 31 | zC_406690 | 2.20 | empty clumpy ring | 25 | 301 | 4.5 | 8.3 | 10.6 | | 0.90 | interacting fluffy satellite | 1.6" W | m: 8:1 | 100? | | |
| 32 | BX610 | 2.21 | extincted bulge & disk | 39 | 327 | 4.9 | 8.2 | 11.0 | | 0.42 | | | | | | |
| 33 | K20_ID7-GS4_29868 | 2.23 | empty clumpy ring | 64 | 308 | 8.2 | 8.2 | 10.3 | | 0.03 | two companions? | 3.2"S 3.1"W | H: 4:1 | | | |
| 34 | K20_ID6-GS3_22466-GS4_33689 | 2.24 | disk | 31 | 162 | 5.0 | 8.2 | 10.4 | | 0.30 | minor merger | 1.3" SW | H: 6:1 | -170 | 0.05 | 0.15 |
| 35 | zC_400569 | 2.24 | bulge & disk | 45 | 289 | 4.0 | 8.2 | 11.1 | | 0.70 | group with 3 members | 1.2" S 2.5" S | m: 11:1 | -280 and -400 | 0.01 | 0.05 |
| 36 | BX482 | 2.26 | small extincted bulge & ring | 60 | 293 | 5.8 | 8.2 | 10.3 | | 0.02 | interacting group with 3 members | 1.9" SW 3.3"SE | H: 10:1 20:1 | 690 and 800 | 0.01 | 0.09 |
| 37 | COS4_02672 | 2.31 | bulge& ring | 62 | 190 | 7.4 | 8.2 | 10.6 | | 0.10 | companion | 2.6" ENE | H 5:1 | | | |
| 38 | D3a_15504 | 2.38 | bulge & disk | 40 | 268 | 6.1 | 8.1 | 11.0 | | 0.30 | interacting satellite | 1.5" NW | m: 30:1 | -50 | 0.06 | 0.08 |
| 39 | D3a_6004 | 2.39 | big bulge & ring | 20 | 416 | 5.3 | 8.1 | 11.5 | | 0.44 | companion? | 1.7" SE | H: 50:1 | | | |
| 40 | GS4_37124 | 2.43 | bulge & red ring/disk | 67 | 260 | 3.2 | 8.1 | 10.6 | 0.20 | 0.70 | | | | | | |
| 41 | GS4_42930 (GK 2363) | 2.45 | bulge & compact disk | 59 | 171 | 2.8 | 8.1 | 10.3 | 0.25 | 0.50 | | | | | | |

Table 1. Overview of the RC41 sample, and its 'input' and environmental properties. We are adopting a $\Omega_m$=0.3, $H_0$=70 km/s/Mpc $\Lambda$CDM Universe, and a Chabrier (2003) initial stellar mass function. Effective radii, inclinations, input stellar masses are derived from analysis of the optical HST ancillary information, using the techniques of Wuyts et al. (2001a, see text). Column 15: $\delta v/v_c(R_e\ (A))=(v_c(R_e(A))/\Delta v)\times(R_e(A)/R_{AB})\times(M_B/M_A)$; column 16: $R_J/R_{AB}$*(de-projected)*$=(R_e(A)/1.5\times R_{AB})\times(M_B/2M_A)^{0.333}$

***Warps.*** Significant m=1,2 and 3 mode, out of plane warps ($\delta z \sim \sin(m(\varphi-\varphi_0))$) are frequently observed in the extended outer HI layers of z~0 disks, including the MW (van der Kruit & Freeman 2011). The MW HI warp starts at $R$~15 kpc, or 4.4 $R_e$(MW) (Bland-Hawthorn & Gerhard 2016), is a superposition of m=0,1 and 2 modes, and reaches an average (maximum) scale height of 2 (3) kpc at R=30 kpc (Levine, Blitz & Heiles 2006). Warps in HI may be fire-hose or buckling instabilities near and outside the (possible) truncation of the stellar disk (Binney & Tremaine 2008, chapter 6, van der Kruit & Freeman 2011). Disk stability analysis shows that such instabilities occur below a critical wavelength of $\lambda < \lambda_{crit}=\sigma^{2}/ G\Sigma_{disk}$, where $\sigma$ is the one-dimensional, in plane velocity dispersion of the disk and $\Sigma_{disk}$ its mass surface density (Toomre 1964, Binney & Tremaine 2008, chapter 6, van der Kruit & Freeman 2011). In hydrostatic equilibrium the z-scale height is given by $h_z = \sigma_z^2 / G\Sigma_{disk}$. This means that the buckling/bending instability is only effective in cold disks with $\sigma_z/\sigma < a_{crit}$ ~0.3-0.6 (Toomre 1964, Araki 1985, Merritt & Sellwood 1994). Since z~1-2.5 SFGs have $\sigma_z/\sigma$~1 (Wisnioski et al. 2015), the buckling instability should be suppressed or less effective. The reflection symmetry of most RC41 RCs excludes strong even modes (m=0,2). Increasing or decreasing rotation velocities on *both* sides of the rotation curve, with equal probability, can occur if the dominant bending mode is odd (m=1 or m=3). Finally, the phase of the warp needs not be aligned with the major axis, and might change with radius. Such precessions should be detectable in the residual velocity maps. Those galaxies that show such residuals, suggesting warps or radial motions, will be discussed in Paper 2.

Summarizing, we find that most of the RCs of the RC41 sample are symmetric in terms of their reflection around the dynamical center. About half of the RCs exhibit a well-defined inner peak and a drop of the rotation curves in the outer disk. Most RCs of the other half are flat, and the small remainder are rising. Only 2 of the 41 galaxies are major mergers. Ten galaxies are demonstrably interacting with a small satellite, or a group, within about 1-3" from the central galaxy, and an additional 5 galaxies have projected small companions that might be satellites. The tidal forces of these small satellites in principle can affect the gas dynamics of the central galaxy in regions closest to the satellite, but we estimate that these perturbations are likely at the 10% level in most of these cases. Furthermore, in none of the cases with such nearby satellites is the overall RC asymmetric. Such perturbations and buckling/bending instabilities can drive warps of the main outer disks. Based on residual velocity maps (Paper 2), deviations from circular rotation in a flat disk do occur in some of our RC41 sample, but are not very large in amplitude, consistent with the fact that high-z disks are kinematically hot and geometrically thick, which in turn is expected to suppress or dampen the buckling and bending instability modes.

### *3.2 Dark Matter Fractions in the Outer Disks of Massive SFGs at High-z*

Table D1 lists the best fit properties derived from our dynamical models of the RC41 sample. Columns 27 and 28 of Table D1 list the dark matter fractions at $R_e$ obtained from the constrained posterior fitting (Appendix A.4) of the entire RC (out to 1.5-3.8 $R_e$), with 1σ uncertainties derived from the MCMC analysis, which is discussed in more detail in Paper 2. Figure 6 shows dark matter fractions, $f_{DM}(R_e)$, as a function of disk circular velocity $v_c(R_e)$ for z=0 disks (upper left panel), and passive, early type galaxies from ATLAS-3D (top right), compared to our RC41 sample in two redshift slices (bottom panels), and to the results of the Illustris-TNG simulation (Lovell et al. 2018).

***Inverse Correlation between Galaxy Mass and Dark Matter Content.*** As is known from studies of low redshift star forming disks, the dark matter fraction at the disk effective radius broadly decreases with increasing stellar, baryonic or halo mass, or circular velocity $v_c(R_e)$ (Courteau et al. 2014 and references therein, Courteau & Dutton 2015). The grey line fit in the upper left panel of Figure 6 ($f_{DM}$=1 - 0.279 × ($v_c$ -50.3 km/s)/100 km/s) shows this well-known inverse relationship. Dwarf galaxies are plausibly dark matter dominated, while the most massive, bulged disks near the Schechter mass have lower dark matter content and approach 'maximal disks' ($f_{DM}$ *(max)* <0.28). The 15 SFGs in the lower redshift slice (z=0.65-1.2) of our RC41 sample, as well as the average of the Wuyts et al. (2016) KMOS$^{3D}$ sample (106 SFGs) in that redshift bin, also mostly fall on this inverse z~0 SFG correlation (lower left panel of Figure 6).

In contrast, very few RC41 SFGs in the higher redshift slice (z=1.2-2.45, lower right panel of Figure 6) are found near the gray line. Seventeen of the twenty-six RC41 galaxies and the Wuyts et al. (2016) average in this redshift bin (92 SFGs) fall across a wide swath of low $f_{DM} \leq f_{DM}$ *(max)*, with no or little dependence on $v_c$. This finding confirms and strengthens the earlier results of Wuyts et al. (2016), Genzel et al. (2017) and Lang et al. (2017), now with an almost 7 times larger galaxy sample.

In the z=1.2-2.45 redshift slice, more than 65% of the RC41 disks are baryon dominated within 1-3 $R_e$, and have dark matter fractions comparable to or less than maximal disks. 11 of 17 SFGs (65%) in this region have $f_{DM} \leq 0.15$, consistent with little or no dark matter, given the typical uncertainties. In contrast, at z=0.65-1.2 only 4 of 15 (27 %) have low dark matter fractions. The work of Courteau & Dutton (2015), Barnabe et al. (2012) and Dutton et al. (2013) indicate that at z=0 the fraction of baryon dominated, massive, bulgy disks, in the same $v_c$- and mass range, is still smaller (<10%). Together these findings suggest that *cosmic time plays a role in setting the central dark matter fractions, in addition to mass or* $v_c$. If correct, this conclusion could explain the inconsistencies between some of the recent studies at z~0.5-3, which we discussed in section 1.3. For instance, Tiley et al. (2019b), Giard et al. (2018) and di Teodoro et al. (2016, 2018) focus on galaxies at z~0.6-1.6 and with log($M_*$/M$_\odot$)~9.5-10.7. They find flat or even rising rotation curves, in excellent agreement with the RC41 in the same part of parameter space. As found in z~0 galaxy studies, the dark matter fraction at one to several $R_e$ is correlated with the shape of the RCs. In RC41 flat or dropping RCs yield modest or low dark matter fractions (0→0.5), while rising rotation curves signal larger dark matter fractions (up to 0.8).

Genzel et al. (2017) already pointed out that the low dark matter content within 1-3 $R_e$ in many of their galaxies means that extrapolation of the dark matter mass to the virial scale with an NFW distribution results in unphysically large baryon to dark matter fractions averaged out to $R_{virial}$: $m_{baryon} > f_{b,c} \sim 0.17$, the cosmic baryon fraction. This suggests that the low dark matter content in the outer disks cannot be extrapolated to the entire halo (Genzel et al., Lang et al. 2017). We will return to this point in section 4.1.

***Comparison of High-z Massive SFGs with z=0 ETGs***. Low dark matter fractions across a wide swath of $v_c$ are characteristic of the z~0 ATLAS-3D passive, early type galaxies (ETGs, upper right panel of Figure 6, Cappellari et al. 2012, 2013, c.f. Genzel et al. 2017). The massive z~2 SFGs are near or above the Schechter mass, $M_{S,*}$~$10^{10.7-10.9}$ $M_\odot$ (e.g. Peng et al. 2010, Ilbert et al. 2010, 2013). They are likely to quench and transition to the ETG population soon after z~2, or merge with another galaxy to form an even more massive ETG (Genel et al. 2008, Conroy et al. 2008). At least a fraction of the ATLAS-3D ETGs could be descendants of the massive z~2 SFGs. *The low dark matter fractions of massive z=0 ETGs could then already be set in their star forming ancestors at z~1.5-2.5*. As a cautionary note, the typically 10-20% dark matter fractions within $R_e$ of the ATLAS-3D ETGs in the upper right panel of Figure 6 do assume a mass-dependent, bottom-heavy IMF (~Chabrier/Kroupa at $v_c$~150 km/s, Salpeter at ~240 km/s, super-Salpeter at ~300 km/s, Cappellari et al. 2012). For a Chabrier/Kroupa IMF, dark matter fractions would go up to 25-35 % for the ETGs in the relevant $v_c$-range (black up-arrow in the upper panels of Figure 6). We note that T. Mendel et al. (in preparation) have observed a sample of massive z~1.4 ETGs as part of the KMOS-Virial survey. They find that essentially all of these ETGs, have <30% dark matter fractions, in agreement with our findings (see Figure 8).

In summary, the RC41 sample confirms and substantially strengthens the findings of Genzel et al. (2017) and Lang et al. (2017) that at z~2 massive SFGs with low or even negligible dark matter fractions within 1-3 $R_e$ are common. With the almost 7 times larger sample presented here compared to Genzel et al. (2017), we can now state that more than 2/3 of the massive SFGs at z=1.2-2.45 are baryon dominated. Given the selection bias of RC41 to large galaxies (Figure 1), this fraction is likely a conservative lower limit. Such low dark matter, large massive SFGs exist also at z=0-1 but they appear to be much rarer there. Similarly low dark matter, baryon dominated passive, early type galaxies at z=0-2 may be descendants of the baryon-dominated z~2 population.

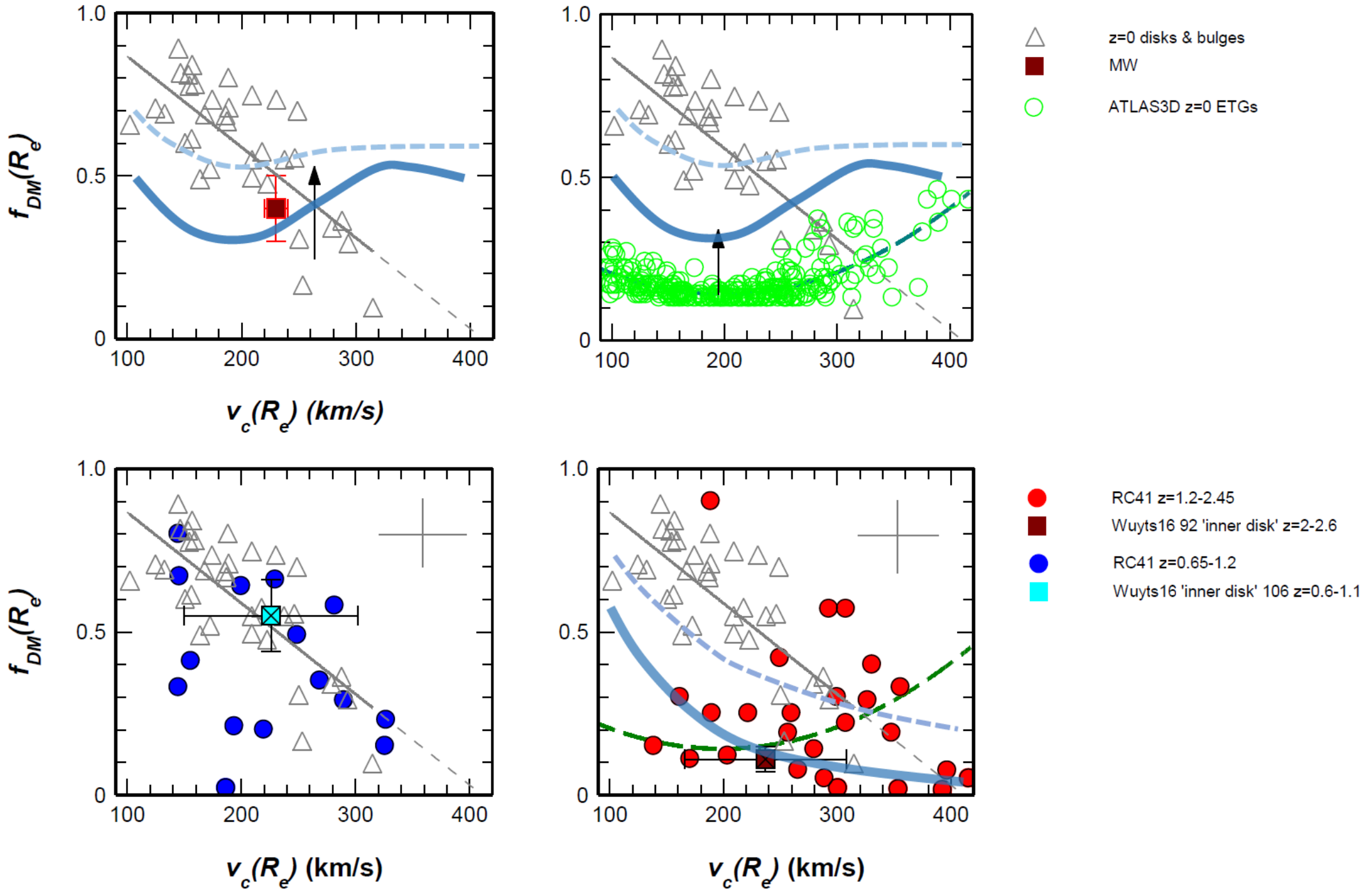

Figure 6. Dark matter fractions within the disk effective radius for star forming disks, as a function of circular velocity at $R_e$ of the disk, and in three redshift bins, z~0 (top left, grey triangles, Martinsson et al. 2013 a, b, Dutton et al. 2013, Barnabe et al. 2012), z=0.65-1.2 (bottom left, blue circles for RC41) and z=1.2-2.45 (bottom right, red circles for RC41). The red/brown square in the upper left panel is the location of the Milky Way (average of Bovy & Rix 2013, and Bland-Hawthorn & Gerhard 2016). The large grey crosses in the bottom two panels denote the median 1σ uncertainties of the RC41 data. The cyan-crossed and brown-crossed square (and ±1σ uncertainties) in the bottom left and right panels are the average dark matter fractions for 106 z=0.6-1.1 SFGs and 92 z=2-2.6 SFGs obtained by Wuyts et al. (2016) from the '*inner disk kinematics*' method. The open green circles in the upper right panel are the early type, passive galaxies from the z=0 ATLAS-3D sample of Cappellari et al. (2012, 2013, 2016), for a bottom-heavy IMF (such as a 'Salpeter' IMF), and the dashed green curve in the top and bottom right panel is the best second order polynomial fit to the ATLAS-3D data. The black up-arrows in the upper panels denote how much galaxies would move upward if instead of a 'Salpeter' IMF a Chabrier (Chabrier 2003), or Kroupa (Kroupa 2001) IMF were chosen. Light blue curves denote the dark matter fractions predicted by Lovell et al. (2018) from the Illlustris-TNG100 simulation, from the full hydro-simulation (dashed) and one with only dark matter (continuous), respectively. The grey line in all four panels ($f_{DM}$=1 - 0.279 × ($v_c$ -50.3)/100) is the best linear fit to the z=0 disks, continued to higher velocity as dashed in the bottom panels.

### *3.3 What Parameters Drive Low Dark Matter Fractions at High-z?*

We now explore more quantitatively, which galaxy properties correlate most strongly with the dark matter content discussed in the last section. Figure 7 shows representative correlation plots. We have also carried out a median-split correlation analysis and a broader Principal Component Analysis, both of which we discuss in Appendix B. The different panels of Figure 7 show $f_{DM}$ *($R_e$)* on the vertical axis, and the respective galaxy property on the horizontal axis. RC41 data are split into z=0.65-1.2 (blue circles, 15 SFGs) and z=1.2-2.45 (red circles, 26 SFGs) redshift slices. The large crosses denote the median uncertainties in these parameters. Best linear fit slopes (and 1σ errors) and the Pearson $r^2$ correlation coefficient are given in red. The thick gray line denotes binned data averages. Figure 7 shows that *dark matter fractions at 1-3 $R_e$ correlate most strongly with baryonic surface density within $R_e$, $\Sigma_{baryon}$ (top right), with baryonic specific angular momentum $\lambda\times(j_{baryon}/j_{DM})$ (top middle), and with bulge mass $M_{bulge}$ (bottom right)*. The correlation with baryonic mass (bottom left), $v_c$, $M_*$ , $M_{DM}$ (bottom center) and galaxy size $R_e$ (top left) are statistically not very significant ($r^2$<0.5). Median-split and PCA analyses give similar results (Appendix B). The PCA analysis also shows that there are no other significant correlations.

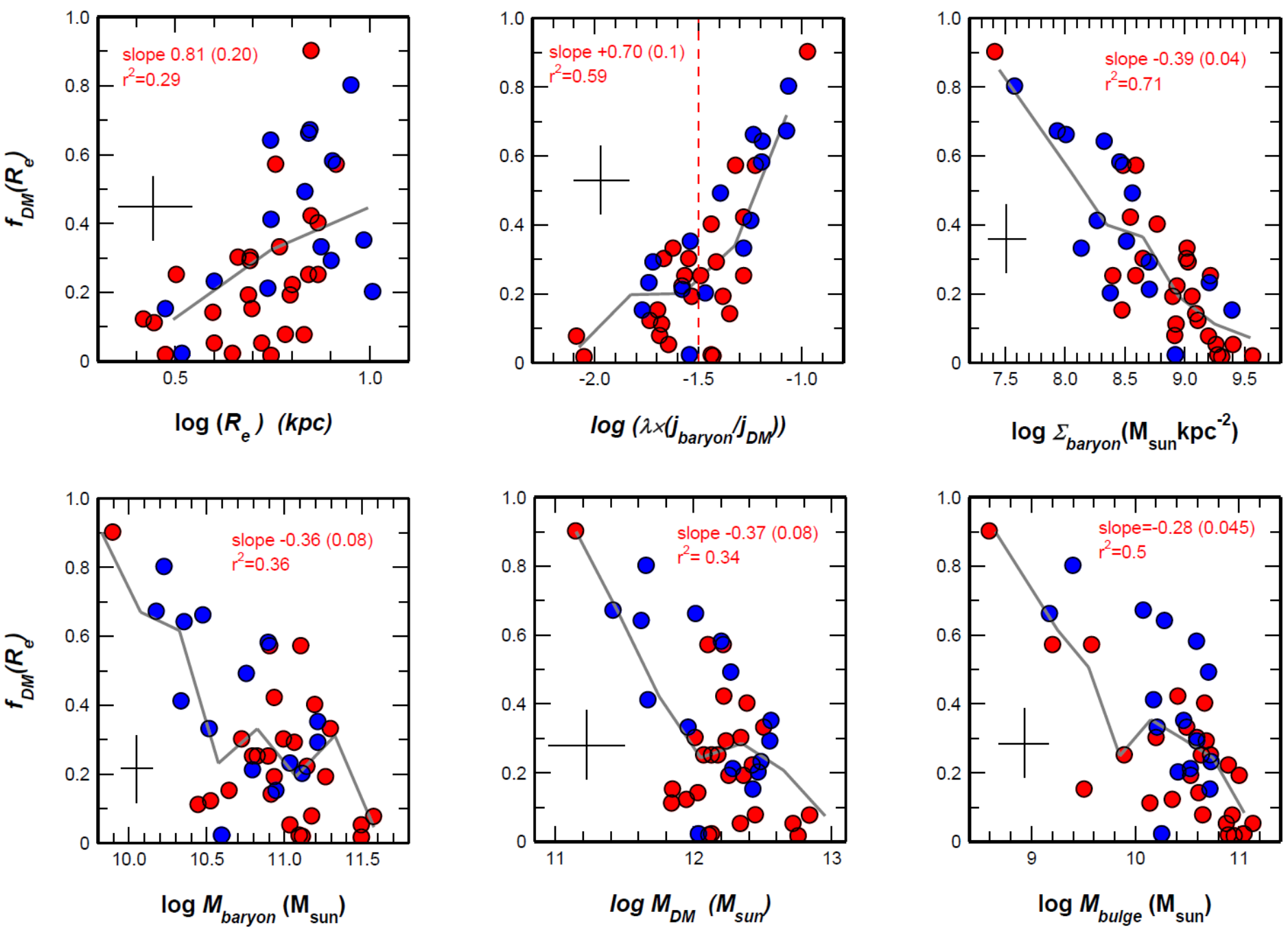

Figure 7. Dark matter fractions of the RC41 sample as a function of basic galaxy parameters. Bottom row: baryonic mass (left), virial masses estimated from the Moster et al. (2018) abundance matching relations (middle, equation A13), and bulge mass estimated from the RC (right). Top row: effective radius of Hα or optical continuum (left), baryonic angular momentum parameter $(j_{baryon}/j_{DM})\times\lambda_{DM}$ (middle, the vertical dashed line indicates the median z=0-2 baryonic and dark matter angular momentum parameter, c.f. Burkert et al. 2016), and baryonic surface density within the effective radius (right). Large black crosses denote the median uncertainties of the parameters in each box. Blue and red circles denote RC41 SFGs in the redshift slices z=0.65-1.2 and 1.2-2.5. The grey curves are the binned running averages. The annotations in red gives the slope (and 1σ uncertainty in brackets) and the Pearson's $r^2$-correlation coefficients of unweighted linear fits for each of the panels. The correlation strength is highest for the $f_{DM}$-$\Sigma_{baryon}$, $f_{DM}$-$M_{bulge}$ and the $f_{DM}$-$(j_{baryon}/j_{DM})\times\lambda_{DM}$ distributions.

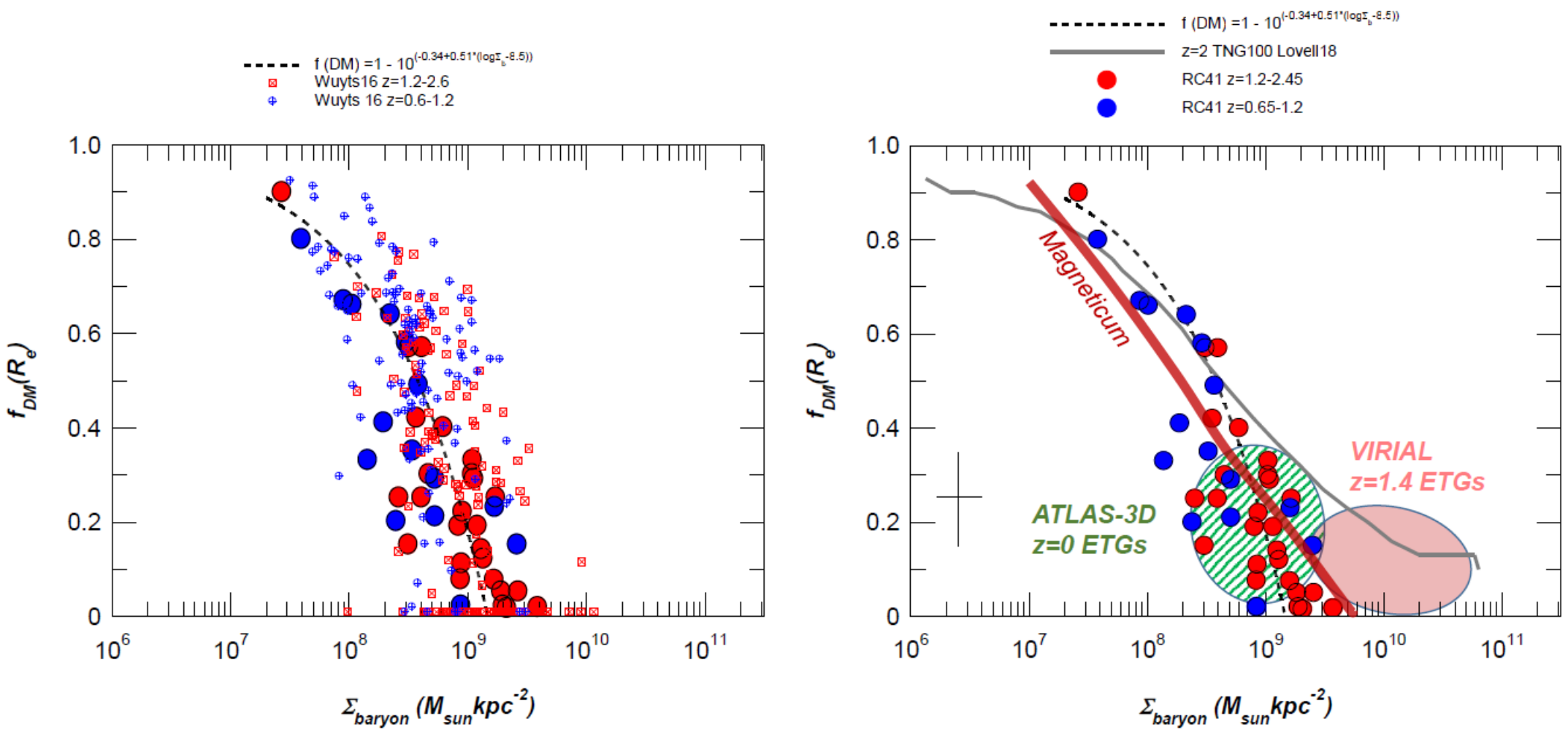

Figure 8. Left: $\Sigma_{baryon}$-$f_{DM}$ correlation for the RC41 sample (same symbols as Figure 7). In addition we include the 240 SFGs of the inner disk kinematics analysis of Wuyts et al. (2016), 106 at z=0.6-1.2 (blue crossed circles) and 134 z=1.2-2.6 (red crossed squares). The Wuyts et al. (2016) sample increases the sample size six-fold and provides an unbiased coverage of the entire MS population, including more compact and higher surface density SFGs than the RC41 sample. The RC41 SFGs and the Wuyts sample show the same trends but the Wuyts et al. (2016) sample has more objects at higher surface density. The dashed grey curve is the correlation between $\Sigma_{baryon}$ and $f_{DM}$ found by Wuyts et al. (2016) (log (1-$f_{DM}$)= -0.34 +0.51 ×(log$\Sigma_{baryon}$-8.5)). Right: $\Sigma_{baryon}$-$f_{DM}$ correlation from RC41 (same symbols as in the left panel, and in Figure 7). The solid grey curve gives the results of TNG100 simulation at z=2 (Lovell et al. 2018, Lovell priv. comm.). The green-shaded ellipse denotes the location of the z=0 ETGs of the ATLAS-3D sample (Cappellari et al. 2012, 2013), and the pink ellipse denotes the locations of z=1.4 ETGs from the KMOS-VIRIAL sample (T. Mendel et al. in preparation).

### *3.4 Extension to the entire z=0.6-2.6 MS Population*

The RC41 sample is biased towards relatively large systems, to ensure well-resolved kinematics with sub-arcsecond resolutions. We are thus missing smaller and denser systems (Figure 1). Given the strong correlation of baryon fraction (= 1 - dark matter fraction) with baryon density in the upper right panel of Figure 7, we expect to find many more low dark matter, baryon-dominated systems in the missing population, consistent with the observations of van Dokkum et al. (2015), Wuyts et al. (2016), Price et al. (2016, 2020) and the simulations of Zolotov et al. (2015).

By including the Wuyts et al. (2016) sample (Appendix C), we can extend the results to a much larger sample that better captures the entire z=0.6-2.6 MS star forming population, and includes smaller, lower mass and higher surface density galaxies. As discussed in the Introduction, Wuyts et al. (2016) have inferred the baryon fractions of 240 SFGs from KMOS$^{3D}$, by comparing the dynamical mass obtained from the rotation curves in the inner disks (to ≤$R_e$) to the sum of stellar and gas masses, thus yielding $f_{baryon}(R_e)=M_{baryon}/M_{dynamical}=1-f_{DM}(R_e)$. Most of the Wuyts et al. (2016) cubes are not as deep as those of RC41, and thus they do not have the extra constraint of RC shape in the outer disk. The dark matter inference thus relies completely on the baryonic mass estimates based on stellar population synthesis modelling of the UV to near-IR SEDs (including IMF, Wuyts et al. 2011a) and the gas scaling relations (Tacconi et al. 2018). Yet the Wuyts et al. (2016) sample provides a check of our results on a six times larger sample. In Figure 8 we show how the RC41 and Wuyts et al. (2016) results compare in the stellar mass - baryonic surface density plane (see upper left panel of Figure 7). Keeping in mind the strong dependence of the Wuyts et al. (2016) dark matter values on systematic uncertainties of the input priors, the agreement of the two samples is remarkable. The Wuyts et al. (2016) sample indeed finds very low dark matter fractions at high surface density, and many more low dark matter, baryon dominated galaxies in the z=1.2-2.6 slice than in the 0.6-1.2 slice. Both samples follow the same strong inversion correlation discussed in the last section. The massive ETGs at z=1.4 (T. Mendel et al., in preparation) and z=0 (Cappellari et al. 2012, 2016) are located in the same region as the massive baryon dominated z~2 SFGs (right panel of Figure 8).

### *3.5 Comparison to Simulations*

The continuous and dashed thick, light blue curves in Figure 6 show the predictions of the Illustris TNG100 simulation at z=0 (top panels) and z=2, with DM profiles taken from a DM-only simulation, and for a full hydro-simulation, respectively (Lovell et al. 2018, Lovell, private communication). Likewise the grey curve in the right panel of Figure 8 shows the z=2 TNG100 full hydro simulation with baryon-dark matter interactions in the $\Sigma_{baryon}$-$f_{DM}$ plane. Broadly the simulations capture the inverse correlation between $f_{DM}$ and $\Sigma_{baryon}$ (Figure 8) and provide a reasonable match to the observed dark matter fractions in z=0 star forming disks/spirals. The simulations also agree with the observational finding that dark matter fractions at a given $v_c$ or $\Sigma_{baryon}$ are greater at lower redshift than at high redshift. In more detail, there are significant differences between the simulations and observations. The baryon-dark matter interaction simulations over-predict dark matter fractions at z~2, especially at the high $v_c$, or high $\Sigma_{baryon}$ tail. In Figure 6 only the pure dark matter simulations come close to the data. This is presumably because the baryon-dark matter interactions pull dark matter efficiently inward, which leads to higher dark matter concentrations at high-z than in a dark matter only model. This is the effect of ‘adiabatic contraction’ (Blumenthal et al. 1986, Mo, Mao & White 1998). The simulations could also miss essential physical processes, presumably on sub-galactic scales, that keep the dark matter fractions as low as in dark matter only simulations. The ‘Magneticum’ simulations (Teklu et al. 2018, Dolag et al., in preparation: thick brown line in the right panel of Figure 8)) come much closer to the observed relation and sharp downturn of dark matter content with surface density. This is because this simulation includes strong AGN feedback.

# 4. Discussion and Interpretation

## *4.1 Physical Meaning of Low Dark Matter Fractions*

Our most important finding of the last section is that *65% (RC41) or more (RC41 +Wuyts et al. 2016) of massive star forming disks at z~2 are baryon dominated within 1-3 $R_e$, and have dark matter fractions comparable to or less than maximal disks ($<f_{DM}> = 0.12 < f_{DM}$ (max) =0.28). At z<1.2 that fraction is 27%, and at z~0 less than 10% of the star forming population appears to have such low dark matter fractions.* Of these baryon dominated, massive ($<M_{baryon}>$ ~$10^{11}$ $M_\odot$) SFGs, 11 of 17 formally have $f_{DM}$ ~ 0→ 0.15, consistent with little or no dark matter, given the typical uncertainties. They are large ($<R_e>$= 5.5 kpc) and have median baryonic angular momentum parameters, log(< $\lambda\times(j_{baryon}/j_{DM})$ >) = -1.4, somewhat ***above*** the median of the overall population (-1.43, Burkert et al. 2016), as expected (Figure 1), but by much less than the rms scatter of the population ($\sigma(\log(\lambda\times(j_{baryon}/j_{DM}))$~0.18). They are ***not*** 'blue nuggets', proposed to be galaxies shortly after a rapid 'compaction event', perhaps triggered by a major merger (Barro et al. 2013, Zolotov et al. 2015). While blue nuggets are more abundant at higher z, they are relatively rare and make up only about 10-15% of the z~2 SFG population in the 10<log($M_*$/$M_\odot$)<11 mass range (Barro et al. 2013).

We have used the fitting functions of Moster et al. (2018, Appendix A3, eq. (A13), see also Behroozi et al. 2010, 2013, Moster et al. 2013, Burkert et al. 2016) as a separate estimate of the RC41 halo masses, given redshift and the best-fit inferred RC41 stellar masses (column 7 of Table D1). The resulting dark matter masses are listed in column 8 of Table D1. We then computed $m_{d,baryon}(Moster)=M_{baryon}/M_{DM}(M_*,z)_{Moster}$ (column 12 of Table D1, filled brown triangles in Figure 9), which is the ratio of the observed baryonic mass in the disk (as obtained from the RC fitting) to the dark matter mass within $R_{virial}$, as inferred from the observed stellar mass and redshift and equation (A13). We also computed $m_{d,baryon}'=M_{baryon}/M_{DM}(NFW,f_{DM}(R_e))$ (column 13 of Table D1, red and blue filled circles in Figure 9 for the usual two redshift slices), which is the ratio of the same observed baryonic mass in the disk, divided by the virial mass of the NFW halo (as obtained from the RC fitting). This halo has a dark matter fraction $f_{DM}$ ($R_e$), which is identical with column 27 in Table D1. The thick horizontal gray line in Figure 9 shows the value that $m_d$ would have if the cosmic baryon fraction, $f_{bc}$=0.17, were all in the disk. Since in reality most of the baryons still reside in the halo, this line represents the upper physical limit of $m_d$. The triangles scatter around $m_{d,baryon}$~0.055, which is typical for the entire z~1-2.5 SFG population (Burkert et al. 2016). Triangles and circles agree with each other for $f_{DM}(R_e)$ ≥ 0.35. For $f_{DM}(R_e)$ < 0.3, however, the extrapolation to the virial radius of an NFW profile with the fitted $f_{DM}(R_e)$ yields too high, or even completely unphysical values for $m_d'$. *This means that the assumption of an NFW dark matter profile on all radial scales cannot be correct in these cases.*

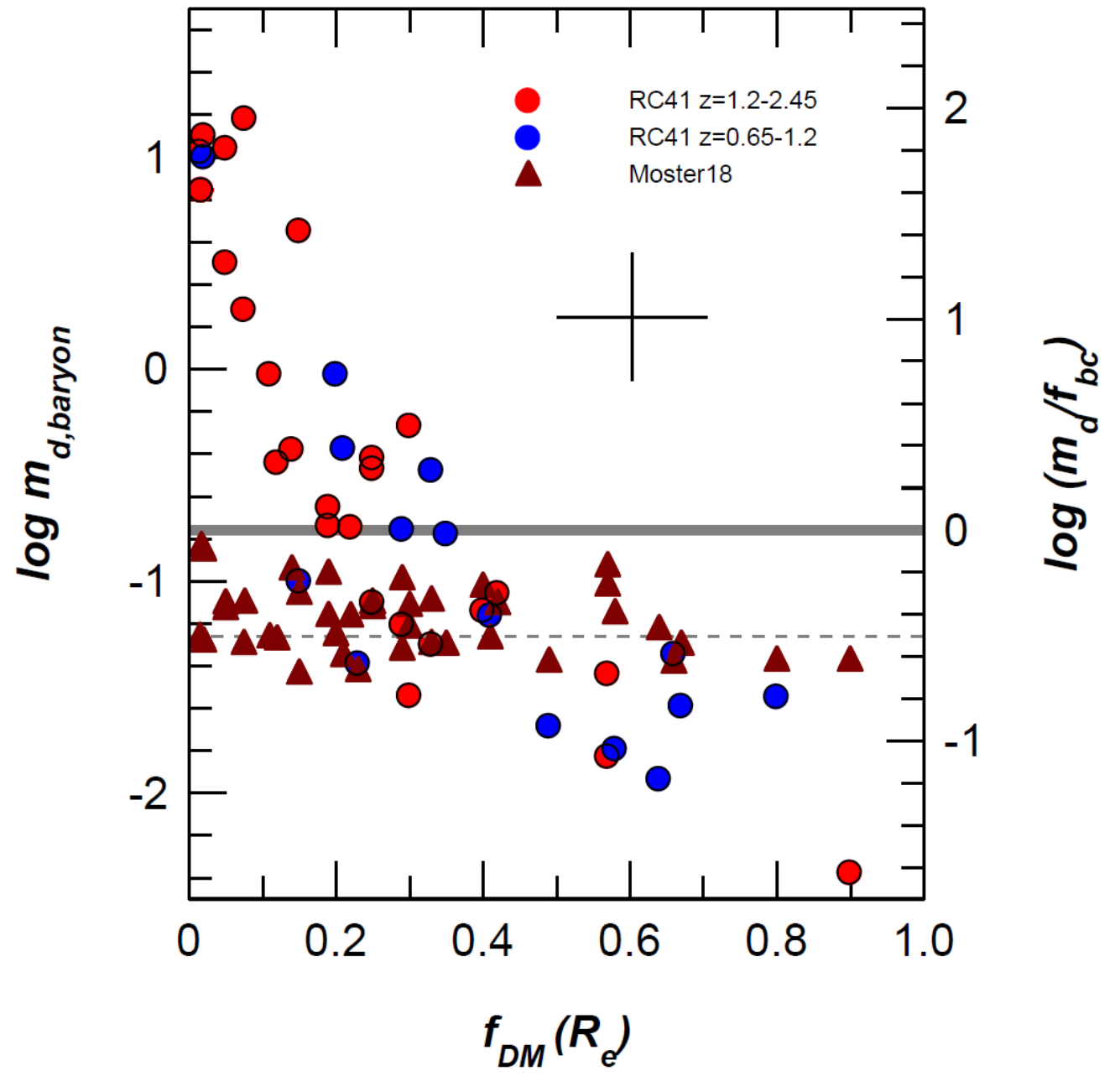

Figure 9. Inferred ratios of total observed baryonic mass in the disk and bulge (including molecular gas) to the total dark matter mass within the halo virial radius (vertical axis), as a function of the inferred dark matter to total mass ratio within $R_e$ of the disk (horizontal axis). Dark brown triangles denote $m_{d,baryon}(Moster)=M_{baryon}/M_{DM}(M_*,z)_{Moster}$, which is the ratio of the baryonic mass in the disk (as obtained from the RC fitting ) to the dark matter mass within $R_{virial}$, as inferred from the best-fit kinematics stellar mass and redshift and equation (A13) from Moster et al. (2018). The blue and red circles denote $m_{d,baryon}'=M_{baryon}/M_{DM}(NFW,f_{DM}(R_e))$, which is the ratio of the same observed baryonic mass in the disk as above, but now divided by the virial mass of the NFW halo (as obtained from the RC fitting). This halo has the dark matter fraction $f_{DM}(R_e)$ inferred from our data (blue for z=0.65-1.2, and red for z=1.2-2.45). The thick horizontal gray line in shows the value that $m_d$ would have if the cosmic baryon fraction, $f_{bc}$=0.17, were all in the disk. Since in reality most of the baryons still reside in the circum-galactic medium, this line represents the absolute upper physical limit of $m_d$. The dotted horizontal grey line marks $m_d$~0.055, the median value in the redshift range z~0.65-2.45 and the mass range of halos considered here (Burkert et al. 2016).

*In our opinion, the most likely cause for the very low dark matter fractions in the majority of our z>1.2 RC41 galaxies are deviations from the NFW profile at small radii.* Flat or cored central dark matter distributions have been inferred for many local dwarf and spiral galaxies (e.g. Flores & Primack 1994, Moore 1994, Burkert 1995, McGaugh & de Blok 1998, de Blok & Bosma 2002, Marchesini et al. 2002, Kuzio de Naray et al. 2006, de Blok et al. 2008, Newman et al. 2012, Faerman, Sternberg & McKee 2013, Oh et al. 2015, Adams et al. 2014), including the Milky Way (Wegg et al. 2016, Portail et al. 2017).

To explore how deviations of the dark matter density distribution, $\rho_{DM}$, from the proposed original profile would affect dark matter fractions, we took equation (A2), but now fitted the RC41 data with a combination of a bulge, a disk and a modified NFW halo, with the additional free fit parameter $\alpha_{inner}$ ($\geq$0), the central slope of the distribution. NFW has $\alpha_{inner}$=1, while a flat core (such as Burkert 1995) has $\alpha_{inner}$=0. We constrain the total mass for the modified NFW distribution by setting the integral of (A2) equal to the inverted Moster et al. (2018) estimate (column 8 in Table D1, eq. (A13)). Columns 29 and 30 of Table D1 give the fit values (and 1 σ uncertainties) for $\alpha_{inner}$.

Figure 10 shows the correlation between $f_{DM}$ and these fitted values of $\alpha_{inner}$. *As expected, very low dark matter fractions are consistent with flat or cored dark matter distributions*. Given the typical virial radii (column 22 in Table D1) and assumed concentration parameters (column 32 in Table D1), the transition to a cored distribution typically would occur at $R_{virial}/c$~24 kpc. Indeed, considering cored Burkert DM distributions (Burkert 1995) with variable $\alpha_{inner}$ and fitting for the core radius $R_{core}$ as the free variable yields $\langle R_{core} \rangle$~25 kpc, consistent with the modified NFW analysis, albeit with large uncertainties. These core radii are also consistent with dark halo core scaling relations (Burkert 2015).

*We conclude that the low dark matter content of the massive high-z SFGs in the outer disks occurs naturally if the central cusp of the NFW profile is replaced by a cored or at least a less cuspy distribution than* $\alpha_{inner}$*=-1.* Replacing the Moster et al. (2018) scaling relations by those of Behroozi et al. (2013), or going from the concentration parameters listed in Table D1 to those one would derive from the Bullock et al. (2001b), Ludlow et al. (2013) or Dutton & Maccio (2014) fitting functions, does not qualitatively change these conclusions.

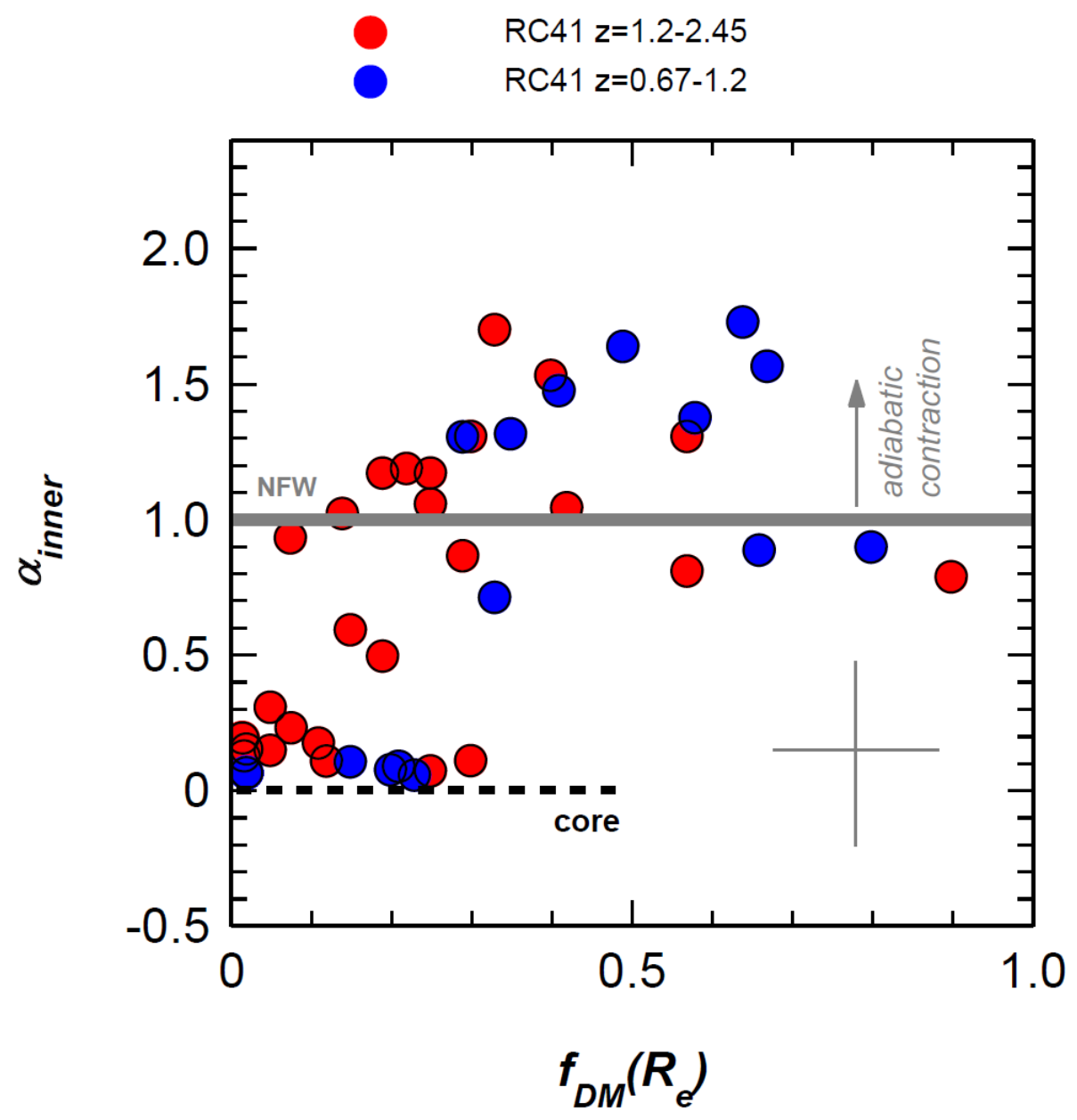

Figure 10. Dependence of the inner slope $\alpha_{inner}$ in a modified NFW-DM distribution (equation (A2)) on the alternative dark matter fraction within $R_e$ for a standard NFW model, as discussed in section 3. A strongly baryon dominated SFG in the former 'language' is synonymous with a cored ($\alpha_{inner}$~0) modified NFW, or similar to a Burkert (1995) distribution.

### *4.3 Possible Drivers of the Dark Matter Deficits/Cores*

Several types of models might explain the presence of dark matter cores. The first explanation considers different dark matter particles, such as warm, fuzzy, or self-interacting dark matter (Hu, Barkana & Gruzinov 2000, Spergel & Steinhardt 2000, Bode, Ostriker & Tork 2001, Bertone, Hooper & Silk 2005, Calabrese & Spergel 2016, Hui et al. 2017, Pozo et al. 2020). The second favors a fundamental change of the law of gravity, MOND (Milgrom 1983). The third type of model considers interactions between baryons and dark matter in the inner halo and circum-galactic medium, where baryonic effects may act rapidly and efficiently (Governato et al. 2010, Maccio et al. 2012, Di Cintio et al. 2014, Chan et al. 2015, Peirani et al. 2017, Teklu et al. 2018, Dolag et al. in preparation).

On balance, a large abundance of baryon dominated, dark matter cored galaxies at z~2, most strongly correlated with baryonic surface density, angular momentum and central bulge mass, may be most naturally accounted for by the interaction of baryons and dark matter during the formation epoch of massive halos. Massive halos (log($M_{halo}/M_\odot$)>12) formed for the first time in large abundances in the redshift range z~1-3 (Press & Schechter 1974, Sheth & Tormen 1999, Mo & White 2002, Springel et al. 2005). At the same time, gas accretion rates were maximal (Tacconi, Genzel & Sternberg 2020). This resulted in high merger rates (Genel et al. 2008, 2009, Fakhouri & Ma 2009), very efficient baryonic angular momentum transport (Dekel, Sari &

Ceverino 2009, Zolotov et al. 2015), formation of globally unstable disks, and radial gas transport by dynamical friction (Noguchi 1999, Immeli et al. 2004, Genzel et al. 2008, Bournaud & Elmegreen 2009, Dekel & Burkert 2014, Bournaud et al. 2014). These processes enabled galaxy mass doubling on a time scale <0.4 Gyrs at z~2-3, and massive bulge formation by disk instabilities and compaction events on < 1 Gyr time scales. However, central baryonic concentrations would naturally also increase central dark matter densities through adiabatic contraction (Barnes & White 1984, Blumenthal et al. 1986, Jesseit, Naab & Burkert 2002). For adiabatic contraction to be ineffective requires the combination of kinetic heating of the central dark matter cusp by dynamical friction from in-streaming baryonic clumps (El-Zant, Shlosman & Hoffman 2001, Goerdt et al. 2010, Cole, Dehnen & Wilkinson 2011), with feedback from winds, supernovae and AGNs driving baryons and dark matter out again (Dekel & Silk 1986, Pontzen & Governato 2012, 2014, Martizzi, Teyssier & Moore 2013, Freundlich et al. 2020, Dolag et al. in preparation). Using idealized Monte Carlo simulations, El-Zant et al. (2001) demonstrated that dynamical friction acting on in-spiraling gas clumps can provide enough energy to heat up the central dark matter component and create a finite dark matter core (see also A. Burkert et al., in preparation). They argue that dark matter core formation in massive galaxies would require that clumps be compact, such that they avoid tidal and ram-pressure disruption, and have masses of $>10^8$ $M_\odot$. Other idealized simulations (e.g. Tonini, Lapi & Salucci 2006) confirm these results.

In full cosmological simulations, repeated and rapid oscillations of the central potential due to supernova and/or AGN feedback result in strong outflows and an expansion of the central collisionless component(s) (Read & Gilmore 2005, Mashchenko, Wadsley & Couchman 2008, Macciò et al. 2012, Pontzen & Governato 2012, Martizzi, Teyssier & Moore 2013, Di Cintio et al. 2014, Chan et al. 2015, Read et al. 2016, Peirani et al. 2017, van der Vlugt & Costa 2019, Dolag et al. in preparation). However, these results are often based on galaxies at specific mass and redshift ranges, and do not generally apply to the massive SFGs at z~0.5-2.5 discussed here. Nevertheless, the presence of giant star-forming clumps and the ubiquity of both stellar feedback-driven and AGN-driven outflows in the population of massive main-sequence galaxies at z~1-3 (e.g. Förster Schreiber et al. 2011, 2019), suggests that the above processes may be at work in RC41 galaxies.

As already mentioned, the results of the currently highest resolution, zoom-in cosmological hydro-simulations (such as Illustris-TNG50/100: Lovell et al. 2018, Pillepich et al. 2019, Nelson et al. 2019) qualitatively agree that high-z massive galaxies have lower dark matter fractions than those at z~0 but do not predict the 'cored' dark matter distributions we infer from RC41 (Figure 6). Future very high spatial resolution cosmological simulations may properly capture dynamical friction and nuclear outflows on sub-cloud scales.

### *4.4 The Dark Matter 'Deficit' on Average is a Fraction of the Bulge Mass*

Assuming for simplicity that the growing virialized dark matter halo plus galaxy system starts with an initial state of a standard NFW dark matter distribution, the 'final', cored state would require that dark matter from the cusp be partially removed to beyond the cusp's core radius. How large is this DM mass for the RC41 sample?

Here we compare again the difference in dark matter mass between the initial pure NFW halo (as estimated from Moster et al. 2018, equation (A13)), and the DM inside $R_e$ for the best-fit $f_{DM}$ NFW-models from section 3. We then plot this 'dark matter' deficit within $R_e$ as a function of bulge mass, since bulge mass is one of the predictors of whether such a deficit (low dark matter fraction) occurs.

The results are listed in column 25 of Table D1 and shown in the left panel of Figure 11. Typical errors (large black cross) are large since we are computing differences of two uncertain numbers. Bulge masses are also more uncertain than baryonic masses. Nevertheless, within these uncertainties the low dark matter fractions of about 2/3 of the RC41 galaxies are consistent with the hypothesis that on average *dark matter equivalent to 30 (±10) % of the final baryonic bulge mass was removed from the galaxy core during bulge formation*.

Another approach is to exploit the fact that central dark matter column densities in NFW models at z~0.5-2.5 are fairly constant, 1 to $3\times10^8$ $M_\odot$ kpc$^{-2}$ (column 23 of Table D1), and are only weakly dependent on $z$, $\lambda$ and $M_*$. In the right panel of Figure 11 we use the N=270 z=0.6-2.6 star forming disks with bulge masses inferred from multi-band HST photometry (Lang et al. 2014) from the 'full' KMOS$^{3D}$/SINS & zC-SINF sample (Burkert et al. 2016, Wisnioski et al. 2015, 2019), and plot them against the average dark matter surface densities within $R_e$ (as estimated from the HST data), obtained from the Moster et al. (2018) scalings (colored distribution). The observed $<\Sigma_{DM}(<R_e)>$ - $M_{bulge}$ data from the RC41 sample are marked as crossed green circles (column 23 of Table D1), clearly demonstrating the deficit compared to the Moster/Behroozi expectations. The filled brown triangles represent $<\Sigma_{DM}(<R_e)> + 0.3\times M_{bulge}/\pi R_e^2$ as a function of $M_{bulge}$. The distribution of the brown triangles is plausibly the same as that of the 'full' sample, or for NFW models.

Further support for a causal connection between the dissipative generation of dark matter cores during the first, gas-rich phase of massive galaxy formation comes from the HST morphologies of the RC41 SFGs. All five z>2, dark matter rich or -dominant SFGs with $f_{DM}(R_e)$=0.4-0.9 are clumpy rings without or with only small central bulges, while the baryon dominated systems have very large bulges, with <B/T>~ 0.5 (column 21 of Table D1). At z<1, only two of seven dark matter rich systems are ring dominated, while the remaining five have substantial bulges.

In 4.1 we noted that the RC41 SFGs are not 'blue nuggets'. This is strictly true in the sense that the blue nuggets *when formed* are compact, while most of the RC41 galaxies have extended disks/rings. However, simulations suggest that after a compaction event a new, long-lived extended gas disk can re-form from inflowing high-angular momentum gas and cold streams, whose inward mass transport is weakened by the presence of the massive central body (Dekel et al. 2020). Some of the RC41 galaxies, especially at z<1.5 may be such galaxies.

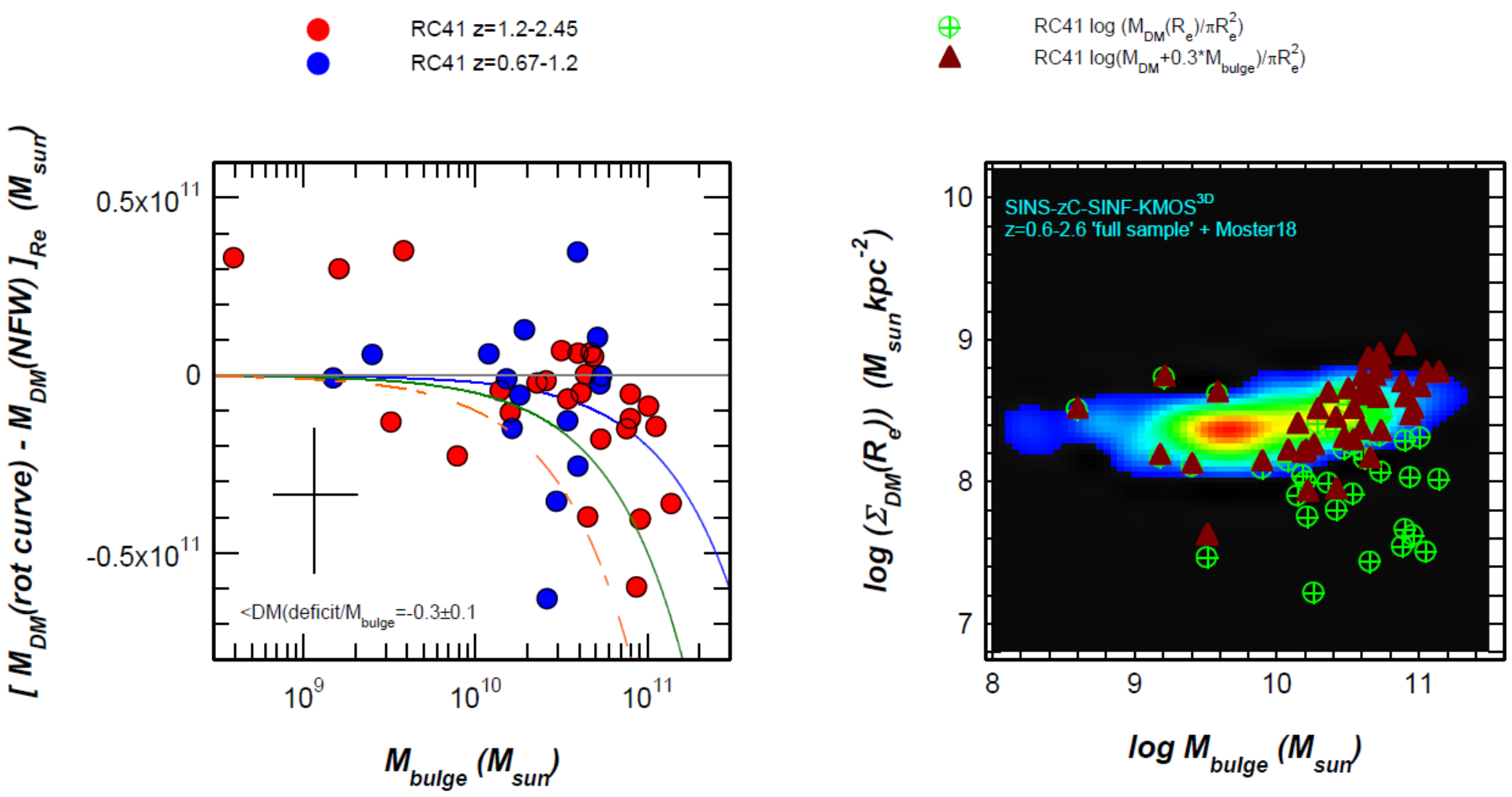

Figure 11. Inferred dark matter ‘deficit’ within $R_e$ as a function of bulge mass for the RC41 sample (columns 24 and 25 in Table D1). Left panel: DM-deficit in two redshift slices (red and blue). The ‘DM deficit’, defined as the difference between the amount of dark matter within $R_e$ inferred from the RC fits and the amount of dark matter expected if the dark matter mass within $R_e$ is computed for a NFW distribution with a virial mass estimated from the Moster et al. (2018) scaling relations between $z$, $M_*$ and $M_{DM}$ (column 8 in Table D1). Blue and red circles denote the RC41 data in the two red-shift slices, as in the other Figures. Negative values of the deficit mean that there is less dark matter actually found compared to the Moster et al. NFW distribution. Blue, green and orange curves denote deficits equal to 0.2, 0.5 and 1 times the bulge mass. Within the uncertainties, the RC41 low dark matter fractions can be explained if dark matter with a mass of 30±10% of the final bulge mass was removed from the core during the formation of the system. Right panel: Another way to present this result is to take N=270 z=0.6-2.6 star forming disks with bulge masses inferred from multi-band HST photometry (Lang et al. 2014) from the ‘full’ KMOS$^{3D}$/SINS & zC-SINF sample (Burkert et al. 2016, Wisnioski et al. 2015, 2019) and plot them against the inferred average dark matter surface densities within $R_e$ (as estimated from the HST data), obtained from the Moster et al. 2018 scalings (colored distribution). The observed $<\Sigma_{DM}(<R_e)>$ - $M_{bulge}$ data from the RC41 sample are marked as crossed green circles, again demonstrating the deficit compared to the Moster et al. expectations from the larger ‘full’ sample. The brown triangles represent $<\Sigma_{DM}(<R_e)> + M_{bulge}/\pi R_e^2$ as a function of $M_{bulge}$.

# 5. Summary

Following from our first deep Hα and CO imaging spectroscopy of outer disk rotation curves in massive z~1-2.5 SFGs (Wuyts et al. 2016, Genzel et al. 2017, Lang et al. 2017, Übler et al. 2017, 2018), we present in this paper RC41, an almost 7 times larger sample of individual rotation curves to 1.5-3.8 $R_e$. RC41 spans a wider range in redshift (z=0.67-2.5) and mass ($\log M_*/M_\odot$=9.5-11.5). Our main results are:

- We confirm the finding in Wuyts et al. (2016), Genzel et al. (2017), and Lang et al. (2017) that at least two-thirds (RC41) of massive star forming disks at the peak of cosmic galaxy/star formation (z>1.2-2.45) had strongly baryon dominated disks on 10-20 kpc scales. Baryon dominance (or low dark matter fraction) correlates most strongly with baryonic angular momentum parameter, baryon surface density and bulge mass. RC41 is biased towards large disks that can be resolved well with current instrumentation. By including smaller galaxies from the Wuyts et al. (2016) KMOS$^{3D}$ sample, the correlation of baryon fraction with baryon surface density fraction of baryon dominated systems at z~1-2 increases still further, in agreement with van Dokkum et al. (2015), Zolotov et al. (2015). These dark matter poor galaxies are likely the ancestors of z=0-1.4 massive ETGs, which also have modest to low dark matter fractions, depending on the IMF adopted;
- The remaining galaxies in the RC41 sample show a wide range of RC shapes, including systems that are strongly dark matter dominated. The fraction of dark matter dominated systems increases toward lower redshift. At z~1 the $v_c$-$f_{DM}$ distribution shows a similar inverse trend (high mass – lower dark matter content) as is well established in the local Universe;
- We show that the low dark matter content of high-z massive disks cannot reflect the entire disk-halo system but most likely relates to shallow or cored central dark matter distributions. The 'dark matter deficit' of these cores compared to classical NFW distributions on average is 30 (±10)% of the bulge mass;
- The strong correlation or anti-correlation of the dark matter content with the angular momentum and surface density of the baryonic disk on 5-20 kpc scales, and the mass of the central bulge on O(1 kpc) scale, suggest that dissipative processes in the gas-rich, early phases of galaxy evolution might be important clues for understanding the central dark matter distributions. The combination of rapid inward transport of baryons, heating by dynamical friction of the dark matter core, and driving out dark matter by the combination of stellar and AGN feedback, can plausibly explain our observations.

***Acknowledgements.*** We are very grateful to the many colleagues at MPE, ESO-Garching and ESO-Paranal, LBT and IRAM, as well as to members of the 3D-HST, SINFONI/SINS & zC-SINF and KMOS/KMOS3D teams, who have contributed to, helped or otherwise supported these observations and their analysis, stretching over more than 14 years on three major optical/IR and Millimeter telescopes. The data analyzed in this paper are based on Hα observations collected with the integral field spectrometers SINFONI (SINS & zC-SINF surveys, plus open time programs) and KMOS (KMOS$^{3D}$ guaranteed time survey) obtained at the Very Large Telescope (VLT) of the European Southern Observatory (ESO), Paranal, Chile (under ESO programmes 073.B-9018,

074.A-9011, 075.A-0466, 076.A-0527, 077.A-0527, 078.A-0600, 079.A-0341, 080.A-0330, 080.A-0339, 080.A-0635, 081.A-0672, 081.B-0568, 082.A-0396, 183.A-0781, 087.A-0081, 088.A-0202, 088.A-0209, 090.A-0516, 091.A-0126, 092.A-0082, 092.A-0091, 093.A-0079, 093.A-0110, 093.A-0233, 094.A-0217, 094.A-0568, 095.A-0047, 096.A-0025, 097.A-0028, 098.A-0045, 099.A-0013, 0100.A-0039, 0100.A-0361, and 0102.B-0087). Next, we included CO observations within the PHIBSS1 and PHIBSS2, open time large projects, and the NOEMA$^{3D}$ guaranteed time project, at the Northern Extended Array for Millimeter Astronomy (NOEMA, located on the Plateau de Bure) Interferometer of the Institute for Radio Astronomy in the Millimeter Range (IRAM), Grenoble, France. IRAM is supported by INSU/CNRS (France), MPG (Germany), and IGN (Spain). Finally, we also include Hα slit-scanning data obtained with the LUCI spectrometer at the Large Binocular Telescope (LBT) on Mount Graham, Arizona, USA. The LBT is an international collaboration among institutions in the United States, Italy, and Germany. LBT Corporation partners are: LBT Beteiligungsgesellschaft, Germany, representing the Max-Planck Society, The Leibniz Institute for Astrophysics Potsdam, and Heidelberg University; The University of Arizona on behalf of the Arizona Board of Regents; Istituto Nazionale di Astrofisica, Italy; The Ohio State University, and The Research Corporation, on behalf of The University of Notre Dame, University of Minnesota and University of Virgina. This work was supported in part by DFG/DIP grant STE/1869 2-1 / GE 625/17-1. Omri Ginzburg and Benjamin Moster kindly provided fitting functions derived from abundance matching, of the relationship between stellar mass and dark matter halo mass. Trevor Mendel kindly gave us access to the location of the z=1.4 KMOS-VIRIAL sample in the $f_{DM}(R_e)$-$\Sigma_{baryon}(R_e)$ plane (Figure 8). We thank Mark Lovell for making his TNG100 results available to us, and Klaus Dolag and Rhea Remus for discussions on the feedback implemented in their Magneticum simulations.

# Appendix A: Data Analysis

## *A.1 Analysis of the RC41 Sample Kinematics*

***Model Assumptions.*** As in our earlier work (Genzel et al. 2006, Wuyts et al. 2016, Burkert et al. 2016, Genzel et al. 2017, Lang et al. 2017, Übler et al. 2017, 2018) we use forward modeling from a parameterized, input mass distribution. This mass distribution is assumed to be combination of a central unresolved bulge $M_{bulge}$ (not radiating in the line emission tracer), a baryonic (gas plus stars) rotating disk $M_{disk}(n_S,R_e,\sigma_{0,},i,pa)$ with Sersic index $n_S$ and disk effective radius $R_e$,[2] inclined by angle $i$ relative to the plane of the sky, and with its major kinematic axis at position angle $pa$ projected on the sky ($pa$ positive east of north). The disk is assumed to have an isotropic, spatially constant velocity dispersion $\sigma_0$ (Wisnioski et al. 2015, Übler et al. 2019). For $n_S$=1 exponential disks characteristic for most of RC41, the RC is given by equation (A3). We assume a flat disk, with a z-scale height determined by hydrostatic equilibrium (Spitzer 1942). We also assume that the disk effective radius $R_e$, which contains half the baryonic mass (gas and stars), is also the disk half-light radius (of the line emission tracer, Hα or CO)[3]. The third component is an extended dark matter halo. The circular velocity of the composite system at radius *R* then is

$$v_{circ}^2(R) = v_{bulge}^2(R) + v_{disk}^2(R) + v_{DM}^2(R) \qquad \text{(A1).}$$

[2] The surface density of this disk is given by $\Sigma = \Sigma_0 \exp(-b_{n_S} \left( R / R_{1/2} \right)^{\frac{1}{n_s}})$ (Burkert et al. 2016, equation (A11)).

For this model we compute from equation A1 the three dimensional circular velocity ($v_c(x,y,z)$) and line intensity ($I(x,y,z)$) distributions, rotate the model given input inclination and position angle of the line of nodes, and compute a 3D 'model cube' $I(x,y,vz)$ after integration of the rotated model distribution along the line of sight. We next convolve with the instrumental response function $PSF(x,y,v_z)$ as appropriate for our measurements.

'***Position-velocity-cuts***'. We discuss the analysis of the RC41 sample in more detail in Paper 2 of this series (S. Price, T. Shimizu, et al., in preparation). As already pointed out in the Introduction, here we focus our analysis on one-dimensional, software slit extractions (slit width *δpp*) of the beam-smeared, projected velocity distribution along the kinematic major axis (position angle *pa*). For these '*position-velocity-cuts*' $I(p, v_z)|_{pa,\delta pp}$ we then compute from Gaussian fits (in apertures of length *δp*~0.15-0.3*"* along the slit) velocity centroid, velocity dispersion and line intensity as a function of *p,* for both the model and measurement cubes. We either use a constant software slit width (typically ~1-1.5 FWHM of the data set), or a fanned slit (with 5-10 degree opening angle), especially for low inclination targets where the expected iso-velocity contours fan out (e.g. van der Kruit & Allen 1978). In cases where OH sky emission lines affect the position velocity cuts modestly, we clip the affected channels and interpolate. In cases of very strong OH sky-line contamination we eliminate the galaxy from the sample. The 1D method is suitable for all RC41 galaxies, since most of the RC information is contained along this major axis (c.f. Genzel et al. 2006, Genzel et al. 2017).

***Dark Matter Model.*** For the dark matter halo, we assume a Navarro, Frenk & White (1996, 1997, henceforth NFW), distribution of total mass $M_{DM}$ (within $R_{200}$, see Mo, Mao & White 1998) and total baryon to dark matter ratio $m_{db}=(M_{bulge}+M_{disk})/M_{DM}$. The *cold dark matter paradigm* (Blumenthal et al. 1984, 1986, Davis et al. 1985) predicts that the dark matter density distribution is described by a universal two parameter analytic function (NFW), alternatively Einasto (1965). The dark matter density distribution in the NFW model is given by

$$\rho_{DM}(R) = \frac{\rho_0}{\left(\dfrac{R}{R_s}\right)^{\alpha_{inner}} \times \left[1+\left(\dfrac{R}{R_s}\right)\right]^{3-\alpha_{inner}}} \qquad \text{(A2),}$$

[3] recent studies at z>0.5 suggests that this assumption is broadly applicable for CO and Hα to first order for star forming, rotating disks (e.g. Figure 5 in Tacconi et al. 2013, Nelson et al. 2016, Wilman et al. 2020). The Hα disks sizes are, however, 10-25% larger than that of the stellar continuum or mass (Nelson et al. 2016, Wilman et al. 2020), whilst the sizes of the submillimeter dust continuum (as tracer of the molecular gas) appear to be substantially smaller than those of the optical tracers (Tadaki et al. 2017, 2020), suggesting that the combination of extinction and radial transport may make more specific statements on sizes dependent on the tracer, the resolution and the properties of individual galaxies.

where $R_s=R_{virial}/c$ is the scale radius of the distribution, and $R_{virial}(z)$ is the virial radius of the DM distribution, within which the mean mass density is ~200 times the critical density of the Universe, and $c$ is the halo's concentration parameter.

Since for dark matter only distributions the dependence of concentration parameter on halo mass is shallow (logarithmic exponent -0.05 to -0.12, Bullock et al. 2001b, Ludlow et al. 2014, Dutton & Maccio at al. 2014), and the range of dark matter halo masses in RC41 is modest, we adopt for simplicity $c\sim10.9\times(1+z)^{-0.83}$ as fixed inputs for our fits (an average of the published work, column 32 in Table D1), without a mass dependence. The scatter between these different theoretical descriptions is about ±0.08 dex.

The NFW distribution is *cuspy* with an inner logarithmic slope of $\alpha_{inner}$=1 (equivalent to $\alpha_{Einasto}$=1.8). At the scale radius $R_s$ the density distribution varies as $R^{-2}$ and in the outer halo as $R^{-3}$. The RC of an NFW distribution has a positive slope from the center to $R_s$, which means that dark matter in the inner core is colder than the gas outside $R_s$. The radial distribution of the dark matter only circular velocity as a function of $R$ is given by

$$v_{DM}^2 = v_{virial}^2 \times \left(\frac{R_{virial}}{R}\right) \times \frac{ln\left(1+R/R_s\right)-\left(R/R_s\right)\left(1+R/R_s\right)}{ln\left(1+c\right)-c/\left(1+c\right)} \quad \text{(A3)},$$

where $v_{virial}$ is the DM circular velocity at $R_{virial}$. The circular velocity reaches a maximum value at 2.2 $R_s$,, remains flat(ish) to about 0.4-0.5 $R_{virial}$, and falls-off beyond. Integration of equation (A2) from $R$=0 to $R_{virial}$ then yields the total dark matter mass $M_{DM}$.

For an infinitesimally thin, exponential disk ($n_S$=1, Freeman 1970) the disk rotation and dark matter circular velocities are given by

$$v_{disk}^2\left(y=R/2R_d\right)=4\pi G\times\Sigma_0 R_d\times y^2\times\left[I_0\left(y\right)K_0\left(y\right)-I_1(y)K_1(y)\right] \quad \text{(A4)},$$

where $I_n$ and $K_n$ denote modified Bessel functions of order $n$.

We assume that the disk is pervaded by a (constant) velocity dispersion $\sigma_0$ (Burkert et al. 2010, 2016), such that the ratio of rotation velocity to dispersion is $v_{disk}/\sigma_0$. If the velocity dispersion is isotropic (Wisnioski et al. 2015) the disk then is geometrically thick ($q=h_z/R_d \sim \sigma_0/v_{disk}$). For this thick disk case, we instead use the rotation curves derived for flattened spheroids of intrinsic axis ratio q=c/a as given by Equation 10 of Noordermeer 2008, solved numerically for a given q and $n_S$. For $v_{disk}/\sigma_0$ ~4 typical at z~2, the disk rotation velocity at $\sim R_{1/2}$ is about 10% smaller than for a Freeman thin disk (Noordermeer 2008). Further, the presence of a pervasive velocity dispersion in an exponential disk creates an outward 'pressure' gradient that decreases the centripetal force. Due to this 'asymmetric drift' the overall rotation velocity becomes (for assumed isotropic and constant velocity dispersion $\sigma_0$)

$$v_{rot}^2 = v_{circ}^2 + 2\sigma_0^2 \times \frac{d\ln\Sigma}{d\ln R} = v_{circ}^2 - 2\sigma_0^2 \times \left(\frac{R}{R_d}\right) \quad \text{(A5)}.$$

The impact of the asymmetric drift correction is two-fold and can be dramatic.

At $R>R_e=1.68\ R_d$ ($n_S$=1) the dark matter mass fraction is $f_{DM}(R)=v_{DM}^2(R)/v_{circ}^2(R)$ and we can rewrite equation (A1), for simplicity for $B/T=M_{bulge}/(M_{bulge}+M_{disk})$~0, as

$$v_{rot}^2(R) \overset{v_{bulge}=0}{=} v_{disk}^2 \times \left[ \frac{1}{1-f_{DM}} - 3.36 \times \left( \frac{\sigma_0}{v_{disk}} \right)^2 \times \left( \frac{R}{R_e} \right) \right] \qquad \text{(A6).}$$

If the disk is not truncated, and if $f_{DM}$ is <1, there is formally a finite radius, at which the rotation velocity becomes small and mathematically even drops to 0 (at the same time the $z$-scale height becomes very large), the system loses the character of a rotation dominated disk and instead becomes a dispersion dominated spherical system. For $f_{DM}\rightarrow 0$ this critical radius is $R_{crit}/R_e \sim 4.8\times(v_{disk}/4\sigma_0)^2$. This drop is much faster than for a baryon dominated but thin Freeman disk (e.g. Figure 2 of Genzel et al. 2017). At or near $R_e$ the asymmetric drift correction, that is the second term within the brackets on the right side of equation (A6), ~ -0.2. This means that if the RC is dropping significantly near and above $R_e$, dark matter fractions have to be small, $f_{DM}(R_e)$<0.2.

To showcase the changes expected ab initio for RCs as a function of redshift, we show in Figure A1 two cases with the same baryonic mass ($M_{baryon}$=8×10$^{10}$M$_\odot$), same $m_{db}$=0.05, same $\lambda\times(j_d/j_{DM})$~0.03, and same $B/T$=0.3. The left galaxy at z=0 is the Milky Way (MW) with c~13, the right galaxy is D3a_15504 at z=2.38 with c=4. The MW model has negligible pressure effects, while D3a_15504 has $\sigma_0$>50 km/s and $v_c/\sigma_0$~4. For the MW case at z=0 the RC is flattish without strong recognizable features of disk and halo. This insensitivity of the RC near $R_e$ to the galaxy parameters is called the ‘disk-halo degeneracy’ (or ‘conspiracy’: van Albada et al. 1985).

At high-z three things happen. First for constant $\lambda\times(j_d/j_{DM})$ $R_e\propto(1+z)^{-1}$ in ΛCDM the baryonic disk is more compact and the peak disk plus rotation velocity correspondingly greater, for constant baryonic mass. Second, because of the smaller concentration of the halo (without considering adiabatic contraction), the halo moves outward and the combined overall rotation curve has a recognizable ‘baryonic peak’ with outer drop. Third, this effect is strongly exacerbated by the pressure term, such that the inferred (and observed) rotation curve drops sharply, indicating the baryonic dominance. The ‘disk halo degeneracy’ *thus can be more easily broken at high-z.*

Mo, Mao & WHite (1998) assume that once a self-gravitating baryonic disk forms, the dark halo contracts adiabatically when the gravitational interactions between baryons and dark matter become important. If so this counteracts the lower halo concentrations and higher baryonic densities of the ab initio model discussed above. However, feedback from supernovae, massive stars and AGN act to expand the halo. Burkert et al. (2010) argue from an analysis of the kinematics of high-z disks that the dark halos did not contract substantially during gas infall and disk formation. We consider both adiabatically contracted (AC) and non-contracted halos in our analysis, and average the resulting dark matter fractions (in columns 27 and 28 of Table D1). The inferred central dark matter fractions are not sensitive to AC for models with both free baryonic and free dark matter masses. For a given best-fit model, however, the effect of AC on the inferred central dark matter fraction is typically ~ +10%.

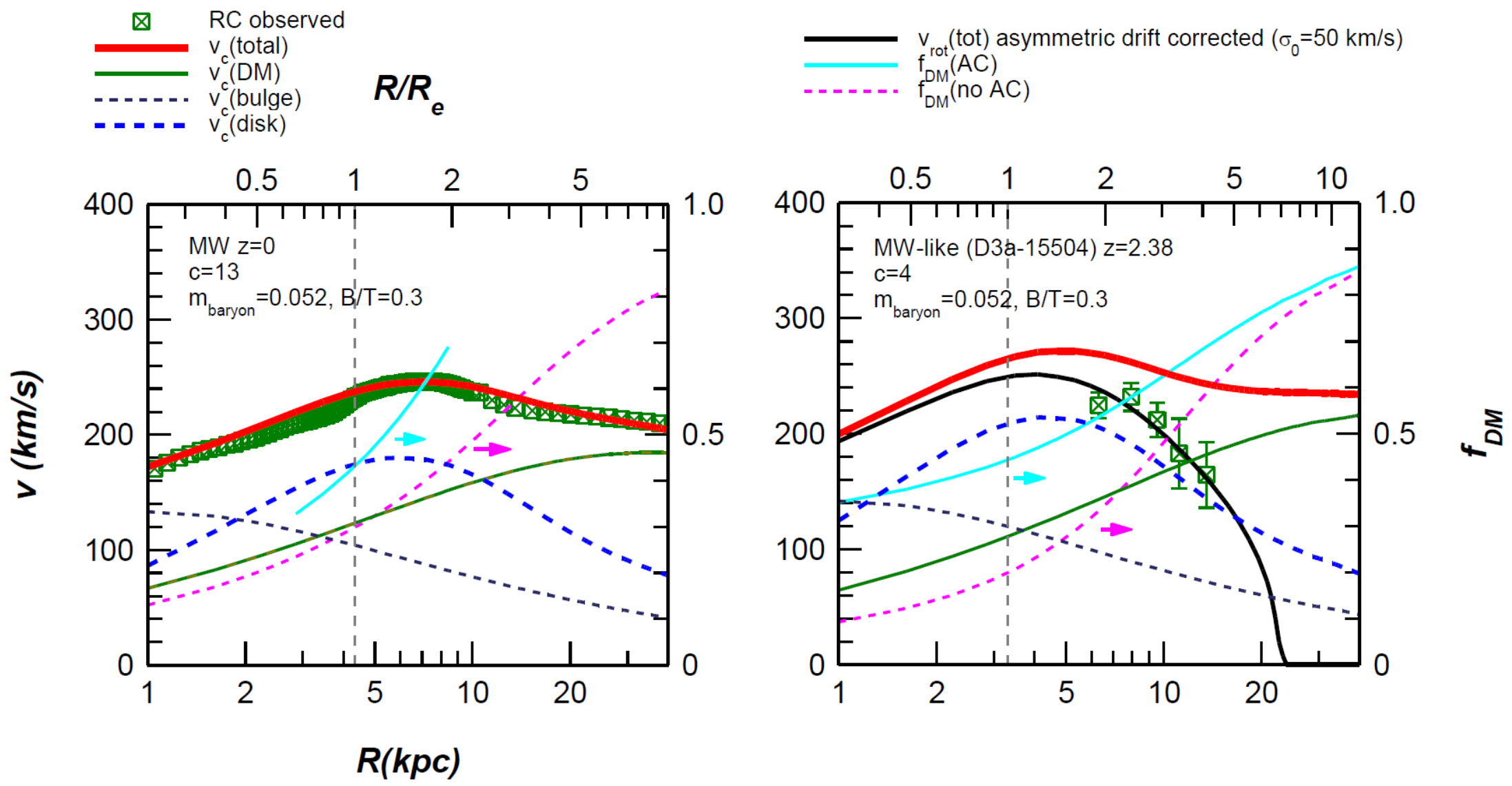

Figure A1. Ab initio expectations of the redshift dependence of RCs, the disk-halo 'degeneracy' and the effects of dispersion ("pressure"). Left panel: total (red curve), disk (dashed blue curve), bulge (dashed black curve) and dark matter (solid green curve) RC curves for a z=0 disk galaxy with the parameters of the Milky Way (left ordinate, Bland-Hawthorne & Gerhard 2016). The observed rotation curve of the MW are the crossed green squares (Sofue 2020, Genzel et al. 2017). The dark matter to total mass fractions without (dashed magenta) and with (solid cyan) adiabatic contraction of the halo (Mo, Mao & White (1998)) and refer to the right ordinate. Right panel: A galaxy with the same mass components in bulge, disk and dark matter, but now at z~2.2, with a halo concentration parameter of c=4 (instead of 13 at z=0), and with 50 km/s velocity dispersion (instead of 10), such that pressure corrections become very important for the rotation curve. As an example of such a galaxy, we show the observed rotation curve of D3a15504 (z=2.38, Figure 2). The color and curve coding for the mass components and observed rotation curve are the same as in the left panel. The additional black curve shows the total RC with the asymmetric drift correction added.

### *A.2 Angular Momentum of Disks*

Tidal torque theory (Peebles 1969, White 1984) suggests that near the virial radius, the centrifugal support of baryons and dark matter is small and given by the angular momentum parameter,

$$\lambda = \frac{\omega_{virial}}{\omega_{virial,cs}} = \varepsilon \frac{J_{DM} / M_{DM}}{R_{virial} \times v_{virial}} \sim \frac{J_{DM} \times E_{DM}^{\ 1/2}}{G M_{DM}^{\ 5/2}} \qquad \text{(A7)},$$

where $\omega = v_{rot}/R$ is the angular speed ($v_{rot}$ is the rotational/tangential velocity) at $R$, and 'virial' and 'cs' stand for 'at the virial radius' and 'centrifugal support' ($\omega_{rot,cs}=(GM/R^3)^{1/2}$). The constant ε is ~√2, $J$ and $j$ are the total and specific ($j=J/M$) angular momenta, and $E{\sim}GM^2/R$ is the absolute value of the total gravitational energy. Building up on earlier work by Peebles (1969) and Barnes & Efstathiou (1987), simulations have shown that tidal torques generate a universal, near-lognormal distribution function of halo angular momentum parameters, with $<\lambda>=0.035\text{-}0.05$ and a dispersion of ±0.2 in the log (Bullock et al. 2001a, Hetznecker & Burkert 2006, Bett et al. 2007, Maccio et al. 2007).

If the baryons are dynamically cold, or can cool rapidly after shock heating at $R_{virial}$, they are transported inwards and form a centrifugally supported disk of (exponential) radial scale length $R_d$, given by (e.g. Fall 1983, Mo, Mao & White 1998)

$$R_d = \frac{1}{\sqrt{2}} \left( \frac{f_{Jd}}{m_d} \right) \times \lambda \times R_{virial} = \frac{1}{\sqrt{2}} \left( \frac{j_d}{j_{DM}} \right) \times \lambda \times R_{virial} \qquad \text{(A8)}.$$

Here $m_d=M_d/M_{DM}$ is the ratio of the baryonic disk mass to that of the dark matter halo and $f_{Jd}$ is the fraction of the total dark halo angular momentum in the disk, $J_d=f_{Jd}\, J_{DM}$. In the classical literature it has generally been assumed that the baryonic angular momentum is conserved between the virial and disk scale, such that $j_d{\sim}j_{DM}$ (e.g. Fall 1983, Mo, Mao & White 1998). More recent simulations find that various competing processes between the outer halo and inner disk scales can lead to up or down variations of $j_d/j_{DM}$ by factors of 3 (Übler et al. 2014, Danovich et al. 2015, Teklu et al. 2015).

***Changes of Angular Momentum due to Turbulent Pressure and Deviations from Exponential Distributions.*** The observations provide an estimate of $\sigma=\sigma_0$ at ~ 2-2.5 $R_{1/2}$, which we adopt as the characteristic dispersion everywhere in the disk (see last section). According to equation (A6), this isothermal disk has a finite 'truncation' radius $R_{max}/R_d = 0.5\times(v_{circ}/\sigma_0)^2{\sim}2\text{-}15$ where $v_{rot} = 0$. For kinematically cold disks with large ratios of rotation-to-dispersion, $v_{rot} = v_{circ}$ and the solution approaches the constant value $R_{1/2}=1.68\times R_d$. Disks with larger velocity dispersions can however be strongly dispersion truncated with half-mass radii that can become even smaller than $R_d$. A convenient approximation is (Burkert et al. 2016)

$$\frac{R_{1/2}}{R_d} = 1.7 \left( \frac{v_{1/2}}{\sigma_0} \right)^{2.7} \times \left( 1 + \left( \frac{v_{1/2}}{\sigma_0} \right)^{2.7} \right)^{-1} \qquad \text{(A9)}.$$

Here, $v_{1/2} = v_{rot}(R_{1/2})$.

Appendix B of Burkert et al. (2016) reformulates (A7) and (A8) in the framework of the Mo, Mao & White (1998) NFW model to yield the following fitting function for the baryonic angular momentum parameter $\lambda\times(j_d/j_{DM})$

$$\lambda = f \times 5.16\cdot10^{-5} \times \left(\frac{j_{DM}}{j_d}\right) \times \left(0.0153c+1\right) \times \left(\frac{R_d \times v_{1/2}}{\text{kpc} \times \text{km/s}}\right) \times \left(\frac{H(z)}{100\ \text{km/s/kpc}}\right)^{1/3} \times \left(\frac{M_{baryon}/m_d}{10^{12} M_\odot}\right)^{-2/3}$$

$$f = 1 \qquad \text{for } \left(v_{1/2}/\sigma_0\right) > 10$$

$$f = 0.5\left[1+\cos\left(\pi\left(\sigma_0/v_{1/2}-0.1\right)\right)\right] \quad \text{for } 10 \geq \left(v_{1/2}/\sigma_0\right) \geq 1.4 \qquad \text{(A10).}$$

$$f = 0.35 \qquad \text{for } \left(v_{1/2}/\sigma_0\right) < 1.4$$

Here, c is the halo concentration parameter, c=$R_c/R_{virial}$ (equation (A2), column 32 of Table D1).

So far we have assumed that the surface density distribution of the baryons is exponential ($n_S$=1). This is motivated by the CANDELS-3D HST analysis of the H-band light of the reference population of the SFGs (Wuyts et al. 2011b, Bell et al. 2012, Lang et al. 2014). 25 of the RC41 SFGs have $n_S$=0.9-1.1. However, 5 of the RC41 galaxies have $n_S$=1.2-1.3 and 10 have $n_S$=0.2-0.6. Romanowsky and Fall (2012) and Burkert et al. (2016) analyzed the combined impact of turbulence (i.e. asymmetric drift) and deviations from exponential distributions on the specific angular momentum of baryonic disks. They cast the results in the form of a fitting function depending on the parameters $x=\log n_S$ and $y=\log(\sigma_0/v_{1/2})$, such that

$$k(n_S, y) = j_d(n_S, y) \big/ \left(v_{1/2} \times R_{1/2}\right), \quad \text{with}$$

$$\Sigma(R, n_S) = \exp\left(-b_{n_S}\left(-\frac{R}{R_{1/2}}\right)^{1/n_S}\right) \text{ and } b_{n_S} = 2n_S - 1/3 + 0.009876/n_S \quad \text{, and}$$

$$\log k(x,y) = -0.082 + 0.091\times x - (0.06 + 0.244\times y)\times x^2 - 0.168\times y \qquad \text{(A11).}$$

Equation (A11) fits all combined data in the interval $x$= -0.7 to 0.7 and $y$= -1.2 to -0.15 to better than ±0.03 dex. For the relevant range in $x$ and $y$, the inferred values of $k(x,y)$ vary from ~1 to ~1.75, where for a thin exponential disk $k(0,-\infty)$=1.19. These corrections tend to slightly decrease $\lambda\times(j_d/j_{DM})$ for SFGs at the low mass tail, and slightly increase $\lambda\times(j_d/j_{DM})$ for SFGs at the high mass end of our sample. These corrected baryonic angular momentum parameters of the disks are listed in Column 26 of Table D1, and used for the top central panel in Figure 7.

### *A.3 Dark Matter Masses from Stellar Masses and Abundance Matching*

Several groups have inferred mean stellar mass to dark matter halo mass ratios as a function of cosmic time and mass, from ranked matching of observed stellar masses of galaxies with computed dark matter masses of the same abundance (both as a function of z and mass, c.f. Behroozi et al. 2010, 2013, Moster et al. 2013, 2018). These analyses give the ratio of $M_*/M_{DM}$, or $M_*$, as a function of $z$ and $M_{DM}$. One can use their fitting functions to in turn compute $M_{DM}$ if $z$ and $M_*$ are given, taking into account the conditional probability distributions. That is, $P(M_{DM}|M_*) = P(M_*|M_{DM})*P(M_{DM})/P(M_*)$, where $P(M_*)= \int dM_{DM} \times P(M_*/M_{DM})$.

Here we use stellar masses either from SED fitting, or inferred from the best-fit baryonic mass together with a fixed gas fraction. For Moster et al. (2018) we have data for individual galaxy-halo pairs at all redshifts and we can directly fit the halo mass-galaxy mass relation. We then find (B. Moster, priv. comm.):

$$\begin{aligned}
&\log M_{DM}(M_\odot) = -0.301 + \log M_* + a + \log(\delta M^b + \delta M^c),\\
&\text{with}\\
&\delta\text{M}=M_*(M_\odot)/3.981\times10^{10}(M_\odot)\\
&a = 1.507 - 0.124\times(z/(1+z)),\\
&b = -0.621 - 0.059\times(z/(1+z)),\\
&c = 1.055 + 0.838\times(z/(1+z)) - 3.083\times(z/(1+z))^2 \qquad \text{(A13).}
\end{aligned}$$

which works well for z>0.5. Using the corresponding relations for Behroozi et al. (2013) gives very similar results. In the RC41 region (log$M_*$~10-11.3 and z=0.65-2.4) we find $\langle \log M_{DM}(\text{M18}) - \log M_{DM}(\text{B13})\rangle_{|z,\log M*}$ =-0.02…-0.1. With $M_{DM}$ in hand it is then possible to compute the dark matter mass distribution $M_{DM}(R)$ and surface density distribution $\Sigma_{DM}(R)=M_{DM}(R)/\pi R^2$ with equations (1) and (A3), using a NFW or modified NFW distribution (i.e. with $\alpha_{inner}<1$).

### *A.4 Fitting Techniques*

***Marquardt-Levenberg gradient $\chi^2$-fitting***. We carried out two independent analyses of the RC41 data set. First, we used a classical Marquardt-Levenberg gradient $\chi^2$ minimization of the $p$-$v(p)$-$\delta v(p)$ and $p$-$\sigma(p)$-$\delta\sigma(p)$ data points at offset $p$ from the center, along the kinematic major axis, and compared to the extraction of our 3D-models at the same positions and smeared to the same angular and velocity resolution. We also used as additional qualitative constraints the $p$-$I(p)$-$\delta I(p)$ measurements of the Hα or CO line but did not include them in most but a few galaxies in the formal fitting since the Hα-light and mass distributions can be quite different in the center if a low-star forming bulge is present. Since the data are sampled every $\delta p$=0.15" to 0.3" we obtain, depending on angular resolution and whether or not adaptive optics was used, between 1.5 to 3.5 samples per FWHM-angular resolution. The total number of data points from the two velocity and velocity dispersion cuts per galaxy then varies between 22 and 60 so that we effectively have 20-30 independent data points (in a Nyquist-sampled sense). To characterize the mass distribution and kinematics, the *minimum* model parameters are $M_{baryon}(disk)=M_{bulge}+(M_{gas}+M_*)_{disk}$, $B/T= M_{bulge}/(M_{gas}+M_*)_{disk}$, $\sigma_0$ and $M_{DM\,(NFW)}$, for a total of 4 free fit parameters. In this case $R_e$, $n_S(disk)$, $i(nclination)_{disk}$, $q_{disk}$, $c_{halo}$, *adiabatic contraction* $AC_{halo}$, are all input parameters constrained/given by ancillary information (mainly HST imaging data, or DM simulations). In some cases, it is necessary to include $R_e$, and in some rare cases also *i* (typically for face-on SFGs) as formal fit parameters, thus increasing the number of fit parameters to 5-6. This leaves 15-25 degrees of freedom. In addition, multi-band photometry analysis gives a prior on $M_*$ (e.g. Wuyts et al. 2011a), and scaling relations give a prior on $M_{gas}$ (e.g. Tacconi et al. 2018, Scoville et al. 2017).

***MCMC Fitting***. Our second approach used a full Bayesian MCMC investigation of the physical parameters. For the present paper, we focus on obtaining uncertainties for the key physical parameters, the dark matter fraction and the total baryonic mass. While the majority of the uncertainty on $f_{DM}$ is captured when fitting only to $f_{DM}$ and log $M_{baryon}$, to mirror the $\chi^2$ analysis we perform MCMC fits with $f_{DM}(R_e)$, log $M_{baryon}$, *B/T*, $\sigma_0$, and if needed, $R_e$ free. The other parameters are fixed to the adopted ancillary parameter values (or the adjusted inclination value). We use flat priors on $f_{DM}(R_e)$ and $\sigma_0$, and Gaussian priors on log $M_{baryon}$, *B/T* and $R_e$, using prior information from ancillary information (but using kinematically derived prior centers for *B/T* and $R_e$, as the light and mass distributions in galaxies can differ). Paper 2 discusses the methodology, results and uncertainties in finding the best fitting values for the basic fit parameters and for the MCMC analysis.

The two methods broadly agree quite well, $<f_{DM}(\chi^2)> - <f_{DM}(MCMC)> = -0.11$ (±0.18), where the number in the bracket is the median 1σ rms uncertainty from the MCMC analysis. The MCMC analysis does give (marginally) higher DM-fractions, mainly for galaxies of low dark matter content in the $\chi^2$-technique. The most likely explanation for the small systematic trend of -0.1 in $f_{DM}$ is the treatment of priors in the two analyses. For the $\chi^2$ analysis we used a top-hat distribution of priors, while in the MCMC analysis we used a Gaussian distribution for the prior on the baryonic mass. In the MCMC fitting there is a trend to favor lower baryonic masses, a number of them more than one unit of dispersion below the prior maximum. This offset then increases the allowable dark matter fraction. The same effect is seen in Wuyts et al. (2016), where their Figures 5, 7 and 9 show a significant number of unphysical values of log $(M_{baryon}/M_{dyn})>0$ (and up to 0.5).

In the following we use the values of $f_{DM}$ obtained from the $\chi^2$-fitting as our final estimators, and adopt the MCMC uncertainties as the more robust estimators of uncertainties (Paper 2).

***Dependence of dark-matter fractions on parameter correlations.*** Figure A2 gives four examples of the dependence of total $\chi^2$ on different galaxy parameters, for the NOEMA CO data set in the PHIBSS1 galaxy EGS_13035123 (z=1.12, Tacconi et al. 2010). These figures show that $\sigma_0$, $R_e$, and *B/T* are constrained to ±15-20%, ±6% and ±10-15% (1σ) in this galaxy. The $\chi^2$-dependence of $f_{DM}(\leq R_e)$ is shallower, and dark matter fractions can rarely be better determined than ±0.1 in absolute units, or ±20-40% in fraction. This fractional uncertainly is increased by at least another ±0.05 if it is unknown whether adiabatic contraction is effective (see discussion in Methods of Genzel et al. 2017).

In terms of covariance, we find that increasing *B/T* leads to increased $f_{DM}(R_e)$, which can be understood in the sense that a higher baryonic mass fraction in the central bulge decreases the relative contribution of the baryons to $v_c$ at $R_e$. When changing $R_e$ the effects are less definite, but for the majority of cases we find that increasing $R_e$ again leads to increased $f_{DM}(R_e)$. For these cases, this result can be understood in the sense that a larger $R_e$ distributes the baryonic mass onto a larger disk (that is, less compact), leading to less relative contribution of the baryons to $v_c$ at $R_e$. When increasing or decreasing the best-fit $R_e$ or *B/T* by their uncertainties given in Figure A2, the resulting changes in $f_{DM}$ are all below ~0.13.

The inferred dark matter fraction depends on the stellar initial mass function (IMF) adopted. We used a Chabrier (2003) IMF for the input stellar priors. A bottom-heavy IMF, as proposed by van Dokkum & Conroy (2010) and Cappellari et al. (2012) for very massive passive galaxies would imply a higher stellar mass for the prior information. A Salpeter IMF (compared to the Chabrier IMF) yields a 0.23 dex greater (M/L*) ratio than the Chabrier IMF. This change in the input prior may push the inferred dark matter fractions to lower values. However, at z~1-2.5 average dense gas to stellar mass ratios in the inner star forming disks of SFGs are ~0.5-1, thus lowering the impact of bottom-heavy IMFs correspondingly. However, our fitting method directly fits for the total baryonic mass, without regard to the form of the IMF. Thus, modulo changes to the priors, assuming a more bottom-heavy IMF would not change the dark matter fractions we find in this study.

The dark-matter fraction depends also on the mass distribution of the halo. If adiabatic contraction (AC) in the inner halo on 1-3 $R_e$ is effective, dark matter fractions go up correspondingly, typically by 10% when using the Mo, Mao & White (1998) AC recipes. Recall our values for $f_{DM}(R_e)$ in columns 27 and 28 of Table D1 are averages between AC on and off.

As mentioned above we adopt a simple scaling of concentration parameter as a function of redshift only, which is an average of published work (Bullock et al. 2001b, Dutton et al. 2014, Ludlow et al. 2014). In reality concentration parameters in simulations are mass dependent (log c~ -0.05*log$M_{DM}$) and show considerable scatter (e.g. Dutton et al. 2014, Ludlow et al. 2014). Lower concentration parameters (at higher z, higher mass etc.) result in somewhat lower $f_{DM}(R_e)$: in most cases, the baryonic mass and effective radius remain largely unchanged over a wide range of assumed concentration parameters, although reducing c (down to c=2) can increase the inferred $M_{baryon}$ by up to ~0.2 dex and increase the effective radius by a few kpc. For instance, at z~2.3 an extreme change from c~4 to c~2 results in $\delta f_{DM}$=-0.05, which can be significant for high mass, low dark matter systems. But generally, varying the concentration parameter does not have a large impact on the resulting fit parameters, including the dark matter fractions. For the galaxy displayed in Figure A2, large changes in c have only a negligible effect on $f_{DM}(R_e)$ and $\chi^2$, as shown in the upper left panel (no adiabatic contraction is considered here). The inset shows the probability distribution function for halo concentration parameters based on the EMERGE model (Moster et al. 2018, 2019) for central haloes at z=1.1 with 10.6<log($M_*$/M$_\odot$)<11.2 and 20<*SFR*<150 [M$_\odot$/yr], where stellar mass and star formation rate include observational uncertainties.

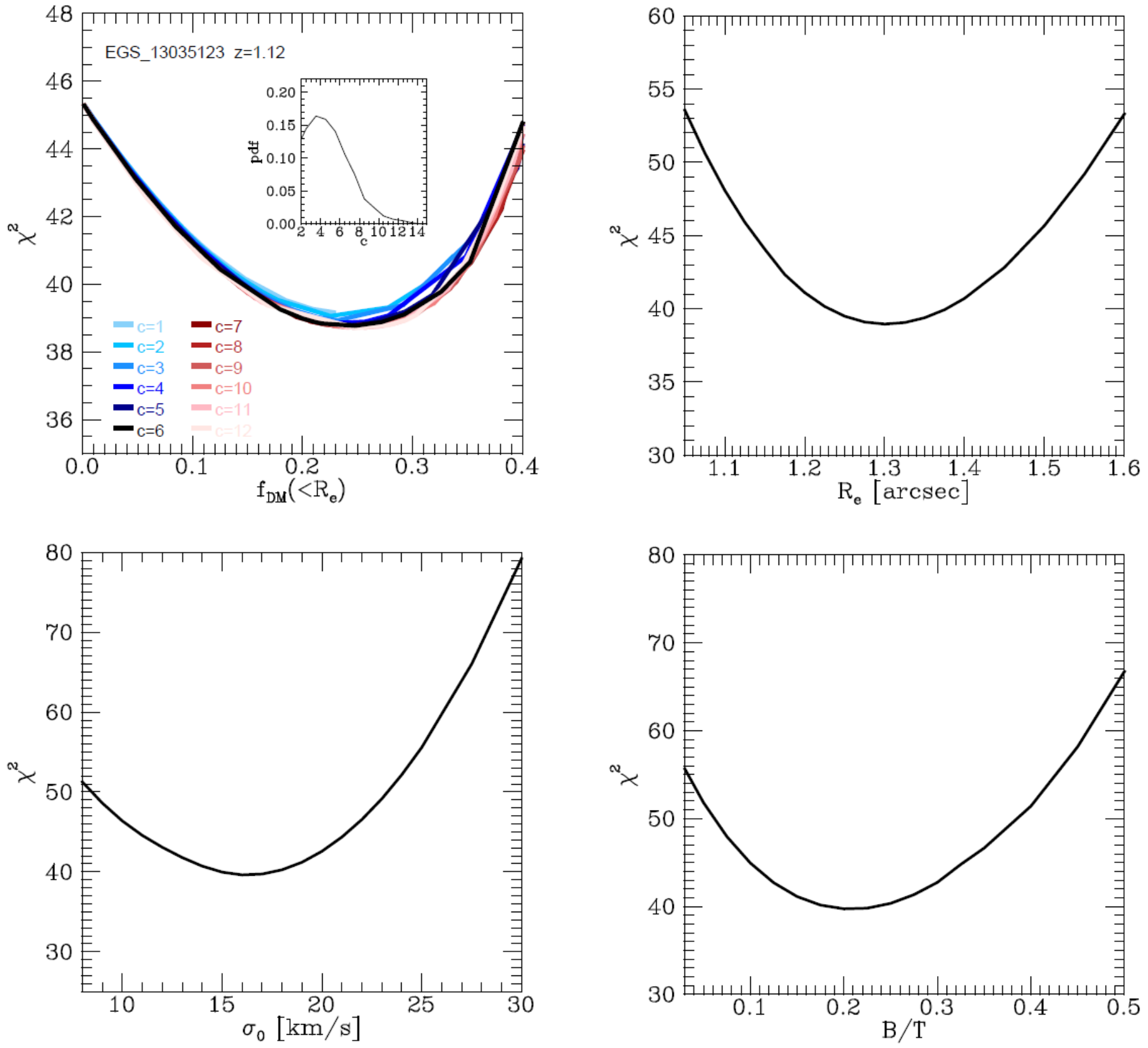

Figure A2. Examples of dependences of the total $\chi^2$ in fits in one of our RC41 galaxies, as a function of 4 parameters, $f_{DM}(\leq R_e)$ (top left), disk effective radius $R_e$ (top right), velocity dispersion $\sigma_0$ (bottom left), and bulge to total baryonic mass ratio *B/T* (bottom right). Fits are performed simultaneously to the one-dimensional velocity, velocity dispersion, and flux profiles. For parameter scans in $R_e$, $\sigma_0$, and *B/T*, the halo mass is a free parameter, while for $f_{DM}(\leq R_e)$, in addition to the scan parameter $M_{DM}$, also $R_e$, $\sigma_0$, and *B/T* are allowed to vary. In the upper left panel we show the effect of different halo concentration parameters from c=1 to c=12 on $f_{DM}(\leq R_e)$ and $\chi^2$ as colored lines, which for this galaxy is negligible. The inset shows the probability distribution function for halo concentration parameters based on the EMERGE model (Moster et al. 2018, 2019) for central haloes at z=1.1 with 10.6<log($M_*$/M$_\odot$)<11.2 and 20<*SFR*<150 [M$_\odot$/yr].

In Paper 2, we investigate the value of using the full 2D velocity and velocity dispersion maps, or, even more ambitiously, fitting the data cubes and full spaxel line profiles. Genzel et al. (2017, Methods section) have shown examples of the 2D-fitting for a few well resolved disks. Paper 2 shows that for most of the RC41 data sets most of the information is contained in the major axis cuts, and additional constraints (e.g. on inclination) from 2D data only are effectively available only for a few (mostly AO-assisted) data sets. Spaxel profile fitting (or parameterizing deviations from Gaussianity with h3 and h4 (Emsellem et al. 2004, Naab & Burkert 2003) can be very helpful for constraining the inner disk/bulge kinematics and structure, and deviations from circular motions. For the outer disk kinematics, which is the focus of the present paper, such an additional analysis effort is not required, since the profiles are close to Gaussian shape in most galaxies.

# Appendix B: Trend Analyses

We used three approaches in this paper to evaluate quantitatively the strengths of the various correlations that naturally occur in a complex galactic system with many, partially degenerate interdependencies. In the first, presented in section 3.3 and Figure 7 of the main text, we explored the correlation strength of linear (or power law) fits between the dark matter fraction and several of the obvious mass and structural parameters of these galaxies. As discussed, the strongest correlations are between $f_{DM}(R_e)$ and $\Sigma_{baryon}(R_e)$, $\lambda \times (j_b/j_{DM})$ and $M_{bulge}$. Correlations with total baryonic or dark matter mass, disk size, or circular velocity are less strong, or not significant.

### *B.1 Principal Component Analysis (PCA)*

Here we quantified trends between pairs of properties with the Pearson correlation coefficient and examined all properties simultaneously through a Principal Component Analysis (PCA). We included as possible vectors the redshift, the masses characterizing each galaxy (the baryonic mass, the dark matter mass at the virial radius and the central bulge mass), the effective disk size and the disk scale angular momentum parameter of the baryons, the potential well depth at the effective radius (i.e. the circular velocity), the baryonic mass surface density within the effective radius, and the inner slope of the DM halo profile $\alpha_{inner}$. We analyzed the RC41 sample alone, and in combination with the Wuyts et al. (2016) "W16" sample (excluding from the latter the 14 galaxies already in RC41). For the RC41+W16 case, the $\alpha_{inner}$ is omitted as it was derived only for the RC41 objects, and the W16 set is restricted to the subset of 133 galaxies that have a stellar bulge mass estimate (while all RC41 galaxies have total stellar+gas bulge masses determined from dynamical modelling).

Figure B1 summarizes the results of the PCA analysis. For the RC41 sample, 72% and 83% of the total cumulated fractional variance are captured by two and three principal components, respectively. For the four times larger, but less well-constrained RC41+W16 sample (174 galaxies), 67% and 79% of the variance are captured by two and three principal components. In the 2-D correlation matrices on the left, we sorted the galaxy parameters from top to bottom going from the strongest correlation to the strongest anti-correlation with $f_{DM}(<R_e)$. Nearly identical results are found using the Spearman rank correlation coefficient instead of the Pearson coefficient. The loading plots from the PCA in the space of the first two principal components (PC1 and PC2), shown in the right-hand panels, reveal that PC1 is strongly coupled to $\Sigma_{baryon}$ and various mass estimates and PC2 is most tightly associated with galaxy size.

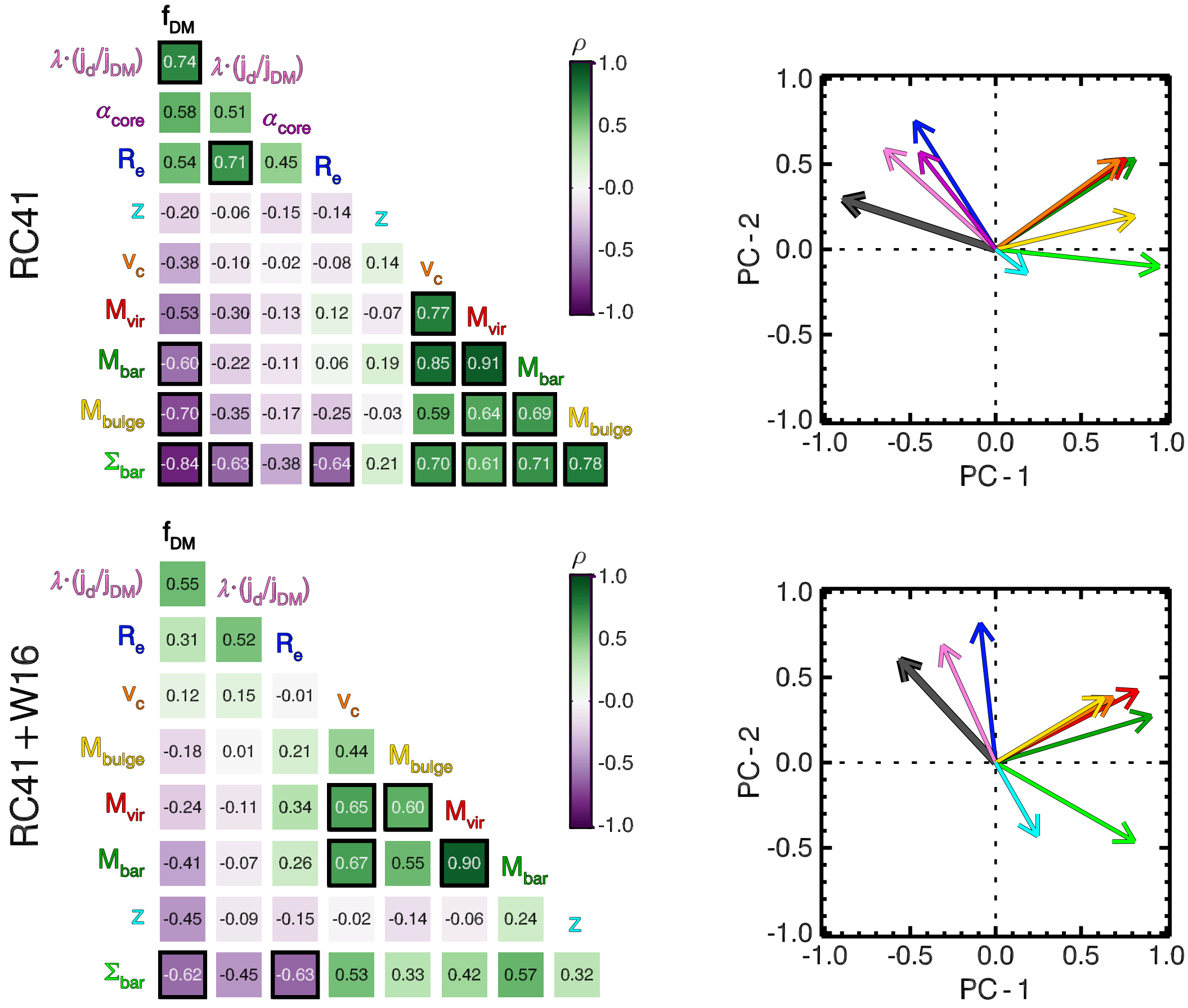

Figure B1. Results of the correlation and PCA analyses for the RC41 sample (top), and the combination of the RC41 and Wuyts et al. (2016) samples (bottom). For each set, we show the correlation matrices of all explored galaxy parameters (left panels), color-coded and sorted from strongest correlation to strongest anti-correlation (following the color bars). The values of the Pearson correlation coefficient are also labeled in the cells of the matrices, and strong to very strong correlations (ρ>0.60) are highlighted with the black outline. The projection of the properties onto the first two principal components (PCs) that account for ~70% of the total sample variance in both data sets are plotted in the right panels. The arrows show the loadings for the different properties, color-coded as the labels in the correlation matrices.

Focusing on $f_{DM}(<R_e)$, the correlation and PCA analyses clearly show that the DM fraction averaged over the disk within $R_e$ is most strongly correlated with $\Sigma_{baryon}$ and $\lambda\times(j_{baryon}/j_{DM})$. For the RC41 sample, we also find a strong correlation with $M_{bulge}$, which becomes much weaker when including the W16 objects perhaps because the bulge masses are based on stellar light measurements from HST data and are not sensitive to gas-rich or highly extincted bulges. The correlations with all other parameters are weak, including redshift, physical size and $v_c$. A more significant trend is seen in RC41+W16 with redshift, presumably because this sample is larger and has many more SFGs in the lower redshift slice, z=0.6-12, than RC41. Expanding the PCA and correlation analysis to all 226 W16 galaxies not in RC41 (for a total of 267) and excluding $M_{bulge}$ has little impact on the results for all other properties, with correlation coefficients changing by 0.05 typically.

### *B.2 Median Splitting Analysis*

We also split the sample into the lower and upper half around the median of each parameter, and then evaluated how significant any difference was in terms of the combined uncertainties of the mean. We carried out this analysis for the RC41 and Wuyts et al. (2016) samples, for the same parameters as in B1. Tables B1 and B2 summarize the results.

| Parameter | value | st-dev | <f_DM> | uncertainty | significance |
|---|---|---|---|---|---|
| **Σ_baryon** | 8.47 | 0.37 | 0.42 | 0.03 | |
| | 9.10 | 0.30 | 0.14 | 0.02 | -7.5 |
| **λ*jd/jDM** | -1.41 | 0.32 | 0.15 | 0.02 | |
| | -1.19 | 0.40 | 0.42 | 0.03 | 7.2 |
| **α_inner** | 0.09 | 0.33 | 0.15 | 0.01 | |
| | 1.10 | 0.40 | 0.42 | 0.03 | 7.1 |
| **R_e** | 4.60 | 0.32 | 0.15 | 0.01 | |
| | 7.10 | 0.44 | 0.34 | 0.03 | 5.3 |
| **M_bulge** | 10.20 | 0.37 | 0.34 | 0.03 | |
| | 10.73 | 0.23 | 0.19 | 0.02 | -3.9 |
| **M_baryon** | 10.63 | 0.28 | 0.32 | 0.02 | |
| | 11.12 | 0.18 | 0.20 | 0.02 | -3.5 |
| **z** | 0.92 | 0.39 | 0.32 | 0.02 | |
| | 2.20 | 0.36 | 0.22 | 0.02 | -2.9 |
| **M_DM** | 11.87 | 0.33 | 0.30 | 0.02 | |
| | 12.30 | 0.22 | 0.22 | 0.02 | -2.6 |
| **v_c** | 194.000 | 0.321 | 0.250 | 0.022 | |
| | 317.000 | 0.199 | 0.225 | 0.027 | -0.7 |

Table B1. Median dark matter fractions in two bins for a given parameter, specified in the $1^{st}$ column. Columns 2 and 3 give the median parameter values (and their standard deviations) in the two bins, the top one being the half of the 41 RC41 galaxies below and the bottom one being the half above the median of a given parameter. Columns 4 and 5 give the median dark matter fractions (and their $1\sigma$ uncertainty) for the two bins. Column 6 gives the significance of the difference between the two bins (upper minus lower), in units of the combined uncertainty. The Table is ordered top to bottom in decreasing level of significance.

| Parameter | value | f_DM (Re) | uncertainty | significance |
|---|---|---|---|---|
| **λ*jb/jDM** | 0.012 | 0.13 | 0.03 | |
| | 0.030 | 0.59 | 0.03 | 11.0 |
| **Σ_baryon** | 8.47 | 0.59 | 0.03 | |
| | 9.06 | 0.24 | 0.03 | -8.2 |
| **z** | 0.91 | 0.55 | 0.03 | |
| | 2.19 | 0.26 | 0.03 | -6.8 |
| **M_baryon** | 10.40 | 0.52 | 0.03 | |
| | 11.00 | 0.29 | 0.03 | -5.4 |
| **M_DM** | 11.69 | 0.49 | 0.03 | |
| | 12.19 | 0.29 | 0.03 | -4.7 |
| **R_e** | 2.85 | 0.32 | 0.03 | |
| | 5.04 | 0.51 | 0.03 | 4.5 |
| **M_bulge** | 9.36 | 0.46 | 0.03 | |
| | 10.32 | 0.29 | 0.03 | -4.0 |
| **v_c** | 184 | 0.38 | 0.03 | |
| | 292 | 0.50 | 0.03 | 2.8 |

Table B2. Same as Table B1, but now for the 240 galaxies of the Wuyts et al. (2016) sample. Note that only 139 SFGs of the Wuyts et al. (2016) sample have stellar bulge masses, estimated from HST imagery. In contrast, for the RC41 sample we estimate the stellar and gas mass content in the bulge region. The Wuyts et al. (2016) bulges thus are lower limits and strongly effected by extinction.

The results in Tables B1 and B2 are in excellent agreement with those obtained in the correlation and PCA analyses in Section B1.

To summarize, we have explored with three methods the significances of the correlation of the dark matter (or baryon) fraction averaged over the disks of z=0.6-2.6 star forming galaxies. All three agree that this quantity empirically is most strongly correlated with baryonic surface density, baryonic angular momentum parameter and, somewhat less convincingly, with the total mass of the central bulge. While these findings may be valuable for understanding the origin of the low dark matter fractions in a surprisingly large number of high mass SFGs at z~2, many of the parameters we have studied, are interrelated and it is difficult, and sometimes misleading, to extract causation from empirical correlations (c.f. Lilly & Carollo 2016).

# Appendix C: The Wuyts et al. (2016) Sample

Wuyts et al. (2016) reported an analysis of the inner disk kinematics of 240 z=0.6-2.6 MS SFGs from the KMOS$^{3D}$@VLT IFS sample (Wisnioski et al. 2015, 2019). This sample consists of massive (log($M_*$/$M_\odot$)>9.8 SFGs with optical sizes $R_e$> 2kpc. The sample was selected from the three CANDELS/3D-HST fields within reach of the VLT: GOODS-South, COSMOS, and UDS (Grogin et al. 2011, Koekemoer et al. 2011, Skelton et al. 2014, Momcheva et al. 2016). Figure C1 shows the distribution of these galaxies in the stellar mass-SFR plane (left), and in the stellar mass–$R_e$ plane. Compared to the RC41 sample (Figure 1) the Wuyts et al. (2016) sample (separated into a z=0.6-1.2, and a 1.2-2.6 redshift slice) not only increases the statistics by another factor 6 but more importantly, it provides an unbiased coverage of the stellar mass–$R_e$ plane, and better coverage of the lower mass tail.

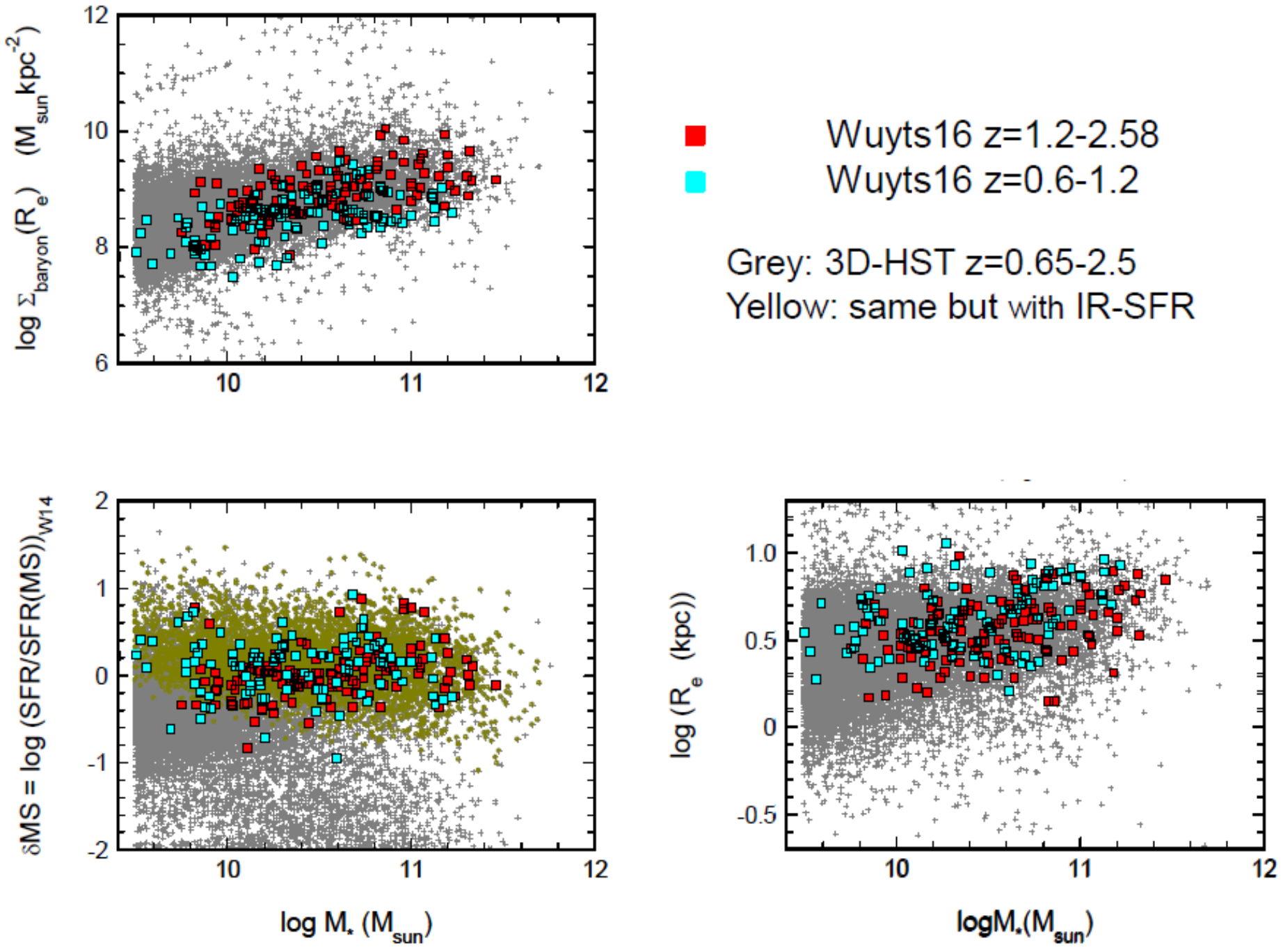

Figure C1. Locations of the Wuyts et al. (2016) SFGs in the planes stellar mass vs. MS-offset (bottom left), stellar-mass vs. effective radius (5000 Å) (bottom right), stellar-mass vs. baryonic surface density (stars plus gas) (upper left). Filled squares denote the location of the 240 galaxies (in redshift bins 0.6-1.2 (cyan), 1.2-2.6 (red)) relative to the galaxies in the 3D-HST parent catalog (Skelton et al. 2014, Momcheva et al. 2016), a near-IR grism survey with the *Hubble Space Telescope* (*HST*) in the CANDELS *HST* imaging survey fields (Grogin et al. 2011, Koekemoer et al. 2011) and with extensive X-ray to radio multi-wavelength data. In the bottom left panel grey crosses denote galaxies for which only SED-based SFRs are available, while yellow crosses denote galaxies where mid- or far-infrared based star formation rates are available in the Wuyts et al. (2011a) 'ladder' scheme of star formation indicators.

# Appendix D

*Table D1*

| 1 | 2 | 7 | 8 | 9 | 10 | 11 | 12 | 13 | 14 | 15 | 16 | 17 | 18 | 19 | 20 |
|---|---|---|---|---|---|---|---|---|---|---|---|---|---|---|---|
| | Target | $\log M_*$ ($M_\odot$) | $\log M_{DM}$ ($M_\odot$) Moster18 | $\log M_{gas}$ ($M_\odot$) | $\log M_{baryon}$ ($M_\odot$) | $D\log M_{baryon}$ ($M_\odot$) fit-input | $m_{d,baryon}$ Moster18 | $m_{d,baryon}$' $f_{DM}$ & NFW | $\delta\log$(SFR/SFR(MS)) W14 | SFR ($M_\odot$/yr) | $v_c$ ($R_e$) (km/s) | $\sigma_0$ (km/s) | $R_e$ (kpc) | $n_S$ | inclination (deg) |
| 1 | EGS3_10098/ EGS4_30084 | 10.85 | 12.39 | 10.24 | 10.95 | -0.26 | 0.0364 | 0.099 | 0.49 | 52.0 | 326 | 15 | 3.0 | 1 | 31 |
| 2 | U3_21388 | 10.41 | 11.87 | 9.66 | 10.48 | -0.35 | 0.0410 | 0.045 | -0.37 | 4.8 | 230 | 55 | 7.0 | 1 | 82 |
| 3 | EGS4_21351 | 10.43 | 11.89 | 10.11 | 10.6 | -0.51 | 0.0518 | 9.953 | 0.72 | 79.5 | 187 | 10 | 3.3 | 1 | 47 |
| 4 | EGS4_11261 | 10.92 | 12.47 | 10.43 | 11.04 | -0.33 | 0.0375 | 0.041 | 0.56 | 85.0 | 327 | 15 | 4.0 | 1 | 60 |
| 5 | GS4_13143 | 10.04 | 11.61 | 10.04 | 10.34 | 0.22 | 0.0534 | 0.068 | 0.33 | 12.0 | 156 | 22 | 5.6 | 1.3 | 74 |
| 6 | U3_05138 | 10.35 | 11.82 | 10.03 | 10.52 | 0.15 | 0.0502 | 0.331 | -0.30 | 6.0 | 145 | 13 | 7.5 | 1 | 55 |
| 7 | GS4_03228 | 9.77 | 11.48 | 9.97 | 10.18 | 0.28 | 0.0499 | 0.026 | 0.48 | 10.1 | 146 | 15 | 7.1 | 1 | 78 |
| 8 | GS4_32796 | 10.54 | 11.99 | 10.36 | 10.76 | 0.17 | 0.0591 | 0.021 | 0.22 | 22.0 | 249 | 40 | 6.8 | 1 | 68 |
| 9 | COS4_01351 | 10.99 | 12.54 | 10.84 | 11.22 | 0.26 | 0.0473 | 0.175 | 0.50 | 57.0 | 290 | 55 | 8.0 | 0.9 | 68 |
| 10 | COS3_22796 | 10.02 | 11.61 | 9.81 | 10.23 | -0.30 | 0.0419 | 0.028 | -0.05 | 11.1 | 145 | 15 | 9.0 | 1 | 58 |
| 11 | U3_15226 | 10.70 | 12.15 | 10.12 | 10.8 | -0.41 | 0.0442 | 0.421 | 0.05 | 31.0 | 194 | 21 | 5.5 | 1 | 50 |
| 12 | GS4_05881 | 9.99 | 11.59 | 10.12 | 10.36 | 0.21 | 0.0591 | 0.012 | 0.46 | 19.0 | 200 | 50 | 5.6 | 1.3 | 60 |
| 13 | COS3_16954 | 10.61 | 12.05 | 10.59 | 10.9 | -0.13 | 0.0701 | 0.016 | 0.55 | 100.0 | 282 | 46 | 8.1 | 1 | 50 |
| 14 | COS3_04796 | 10.99 | 12.52 | 10.83 | 11.22 | 0.19 | 0.0500 | 0.166 | 0.25 | 51.0 | 269 | 10 | 9.7 | 1.07 | 50 |
| 15 | EGS_13035123 | 10.90 | 12.38 | 10.73 | 11.12 | -0.28 | 0.0555 | 0.942 | 0.31 | 126.0 | 220 | 15 | 10.2 | 1 | 24 |
| 16 | EGS_13004291 | 10.51 | 11.96 | 11.00 | 11.12 | -0.46 | 0.1455 | 6.938 | 1.25 | 630.0 | 354 | 50 | 3.0 | 1.3 | 27 |
| 17 | EGS_13003805 | 11.20 | 12.76 | 11.19 | 11.5 | -0.03 | 0.0545 | 10.541 | 0.56 | 200.0 | 393 | 30 | 5.6 | 1.2 | 37 |
| 18 | EGS4_38153 | 10.94 | 12.39 | 11.05 | 11.3 | 0.50 | 0.0807 | 0.050 | 0.25 | 78.0 | 356 | 68 | 5.9 | 1 | 75 |
| 19 | EGS4_24985 | 10.79 | 12.22 | 10.58 | 11 | -0.11 | 0.0600 | 0.029 | 0.37 | 99.0 | 300 | 30 | 4.6 | 1 | 40 |
| 20 | zC_403741 | 10.32 | 11.81 | 10.12 | 10.53 | -0.33 | 0.0529 | 0.360 | 0.30 | 60.0 | 204 | 64 | 2.6 | 1 | 28 |
| 21 | D3a_6397 | 10.89 | 12.32 | 10.81 | 11.15 | -0.19 | 0.0678 | 0.179 | 0.53 | 214.0 | 308 | 67 | 6.4 | 1 | 30 |
| 22 | EGS_13011166 | 10.81 | 12.24 | 11.08 | 11.27 | -0.27 | 0.1082 | 0.222 | 0.74 | 375.0 | 348 | 40 | 6.3 | 1 | 60 |
| 23 | GS4_43501 GK_2438 | 10.69 | 12.11 | 10.59 | 10.94 | 0.08 | 0.0680 | 0.181 | 0.07 | 53.0 | 257 | 39 | 4.9 | 0.6 | 62 |
| 24 | GS4_14152 | 11.38 | 12.88 | 11.15 | 11.58 | 0.08 | 0.0504 | 1.901 | 0.13 | 167.0 | 397 | 35 | 6.8 | 1 | 55 |
| 25 | K20_ID9- GS4_27404 | 10.63 | 12.05 | 10.64 | 10.94 | -0.02 | 0.0776 | 0.087 | 0.03 | 81.0 | 250 | 18 | 7.1 | 1 | 48 |
| 26 | zC_405501 | 10.14 | 11.71 | 10.49 | 10.65 | 0.22 | 0.0868 | 4.467 | 0.60 | 60.0 | 139 | 60 | 5.0 | 0.2 | 75 |
| 27 | SSA22_MD41 | 9.20 | 11.28 | 9.80 | 9.9 | -0.66 | 0.0420 | 0.004 | 0.60 | 130.0 | 189 | 71 | 7.1 | 0.4 | 72 |
| 28 | Q2343- BX_389 | 10.86 | 12.23 | 10.94 | 11.2 | 0.26 | 0.0936 | 0.072 | 0.10 | 100.0 | 331 | 76 | 7.4 | 0.2 | 76 |
| 29 | zC_407302 | 10.38 | 11.87 | 10.77 | 10.92 | -0.01 | 0.1131 | 0.416 | 0.77 | 340.0 | 280 | 22 | 4.0 | 1 | 60 |
| 30 | GS3_24273 | 10.51 | 11.96 | 10.54 | 10.83 | -0.49 | 0.0744 | 0.338 | 0.31 | 267.0 | 222 | 20 | 7.0 | 1 | 60 |
| 31 | zC_406690 | 10.51 | 11.96 | 10.97 | 11.1 | -0.11 | 0.1389 | 12.589 | 0.55 | 300.0 | 301 | 68 | 4.5 | 0.2 | 25 |
| 32 | Q2343- BX_610 | 10.65 | 12.06 | 10.86 | 11.07 | -0.35 | 0.1016 | 0.062 | -0.34 | 60.0 | 327 | 65 | 4.9 | 1 | 39 |
| 33 | K20_ID7- GS4_29868 | 10.63 | 12.04 | 10.94 | 11.11 | 0.35 | 0.1170 | 0.036 | 0.30 | 101.0 | 308 | 68 | 8.2 | 0.2 | 64 |
| 34 | K20_ID6- GS3_22466- GS4_33689 | 10.36 | 11.85 | 10.49 | 10.73 | -0.07 | 0.0760 | 0.537 | 0.17 | 99.0 | 162 | 60 | 5.0 | 0.5 | 31 |
| 35 | zC_400569 | 10.79 | 12.17 | 10.68 | 11.04 | -0.29 | 0.0737 | 10.965 | 0.19 | 240.0 | 289 | 45 | 4.0 | 1 | 45 |
| 36 | Q2346- BX_482 | 10.48 | 11.93 | 10.71 | 10.91 | 0.22 | 0.0948 | 0.015 | 0.20 | 80.0 | 293 | 67 | 5.8 | 0.2 | 60 |
| 37 | COS4_02672 | 10.58 | 12.00 | 10.62 | 10.9 | 0.01 | 0.0788 | 0.378 | -0.10 | 72.0 | 190 | 55 | 7.4 | 0.5 | 62 |
| 38 | D3a_15504 | 10.94 | 12.28 | 10.81 | 11.18 | -0.10 | 0.0789 | 15.136 | -0.06 | 146.0 | 268 | 68 | 6.1 | 1 | 40 |
| 39 | D3a_6004 | 11.30 | 12.61 | 11.07 | 11.5 | -0.20 | 0.0783 | 3.162 | -0.05 | 355.0 | 416 | 68 | 5.3 | 0.4 | 20 |
| 40 | GS4_37124 | 10.44 | 11.91 | 10.55 | 10.8 | -0.15 | 0.0781 | 0.079 | 0.28 | 194.0 | 260 | 50 | 3.2 | 1 | 67 |
| 41 | GS4_42930 GK_2363 | 10.13 | 11.72 | 10.17 | 10.45 | -0.20 | 0.0541 | 0.939 | -0.09 | 70.0 | 171 | 40 | 2.8 | 1.2 | 59 |

| 1 | 2 | 21 | 22 | 23 | 24 | 25 | 26 | 27 | 28 | 29 | 30 | 31 | 32 | 33 | 34 | 35 |
|---|---|---|---|---|---|---|---|---|---|---|---|---|---|---|---|---|
| | Target | B/T | $R_{virial}$ (kpc) Moster18 | $S_{DM}(R_e)$ ($M_\odot$ kpc$^{-2}$) Moster18 | $S_{DM}(R_e)$ ($M_\odot$ kpc$^{-2}$) from fit | $DM_{DM}(R_e)$- deficit data- Moster18 | $\lambda'$($j_{baryon}/j_{DM}$) thick, turbulent disk | $f_{DM}$ ($R_e$) | $Df_{DM}$ | $\alpha_{inner}$ | $D\alpha_{inner}$ | $R_{out}/R_e$ | c DM-halo | Mode | FWHM (arcsec) | $T_{int}$ (h) |
| 1 | EGS3_10098/ EGS4_30084 | 0.60 | 218 | 3.1E+08 | 2.1E+08 | -2.7E+09 | 0.0262 | 0.15 | 0.143 | 0.10 | 0.32 | 2.5 | 7.3 | NOEMA | 0.75" | 21 |
| 2 | U3_21388 | 0.05 | 145 | 1.5E+08 | 1.5E+08 | -9.9E+08 | 0.0621 | 0.66 | 0.093 | 0.88 | 0.22 | 2.0 | 7.2 | KMOS | 0.54" | 8 |
| 3 | EGS4_21351 | 0.46 | 144 | 1.9E+08 | 1.6E+07 | -5.9E+09 | 0.0307 | 0.02 | 0.120 | 0.06 | 0.18 | 3.5 | 6.8 | NOEMA | 0.76" | 19 |
| 4 | EGS4_11261 | 0.50 | 223 | 3.0E+08 | 2.9E+08 | -4.4E+08 | 0.0322 | 0.23 | 0.100 | 0.05 | 0.15 | 3.1 | 6.6 | NOEMA | 0.86" | 24 |
| 5 | GS4_13143 | 0.70 | 115 | 1.2E+08 | 1.1E+08 | -1.3E+09 | 0.0558 | 0.41 | 0.105 | 1.47 | 0.30 | 2.9 | 6.5 | KMOS | 0.52" | 9 |
| 6 | U3_05138 | 0.50 | 133 | 1.4E+08 | 5.6E+07 | -1.5E+10 | 0.0552 | 0.33 | 0.139 | 0.71 | 0.42 | 1.5 | 6.5 | SINF-J | 0.73" | 16 |
| 7 | GS4_03228 | 0.80 | 102 | 1.0E+08 | 1.4E+08 | 5.8E+09 | 0.0766 | 0.67 | 0.092 | 1.56 | 0.22 | 1.5 | 6.5 | KMOS | 0.53" | 8 |
| 8 | GS4_32796 | 0.90 | 150 | 1.8E+08 | 2.8E+08 | 1.5E+10 | 0.0615 | 0.49 | 0.072 | 1.63 | 0.19 | 2.2 | 6.5 | KMOS | 0.52" | 7 |
| 9 | COS4_01351 | 0.24 | 227 | 3.0E+08 | 1.7E+08 | -2.6E+10 | 0.0410 | 0.29 | 0.136 | 1.30 | 0.28 | 2.2 | 6.5 | SINF-J +KMOS | 0.64" | 12 |
| 10 | COS3_22796 | 0.15 | 108 | 1.1E+08 | 1.3E+08 | 5.6E+09 | 0.0861 | 0.80 | 0.168 | 0.89 | 0.39 | 1.7 | 6.1 | SINF-J | 0.65" | 12 |
| 11 | U3_15226 | 0.55 | 164 | 2.2E+08 | 8.2E+07 | -1.3E+10 | 0.0335 | 0.21 | 0.084 | 0.09 | 0.23 | 3.1 | 6.1 | KMOS | 0.75" | 20 |
| 12 | GS4_05881 | 0.85 | 103 | 1.3E+08 | 2.6E+08 | 1.3E+10 | 0.0645 | 0.64 | 0.055 | 1.72 | 0.15 | 2.4 | 6 | KMOS | 0.57" | 14 |
| 13 | COS3_16954 | 0.50 | 146 | 1.9E+08 | 3.6E+08 | 3.4E+10 | 0.0763 | 0.58 | 0.076 | 1.37 | 0.30 | 2.2 | 6.1 | SINF-J | 0.9" | 4 |
| 14 | COS3_04796 | 0.18 | 208 | 2.9E+08 | 1.7E+08 | -3.6E+10 | 0.0649 | 0.35 | 0.132 | 1.31 | 0.37 | 1.6 | 6.1 | SINF-J | 0.7" | 10 |
| 15 | EGS_13035123 | 0.20 | 180 | 2.6E+08 | 6.3E+07 | -6.3E+10 | 0.0594 | 0.20 | 0.100 | 0.07 | 0.21 | 1.8 | 6 | NOEMA | 0.6" | 56 |
| 16 | EGS_13004291 | 0.61 | 127 | 2.4E+08 | 4.6E+07 | -5.6E+09 | 0.0434 | 0.02 | 0.083 | 0.12 | 0.44 | 2.3 | 6 | NOEMA | 0.5" | 38 |
| 17 | EGS_13003805 | 0.29 | 233 | 4.6E+08 | 4.1E+07 | -4.1E+10 | 0.0388 | 0.02 | 0.104 | 0.19 | 0.44 | 2.1 | 5.9 | NOEMA | 0.63" | 20 |
| 18 | EGS4_38153 | 0.16 | 167 | 2.9E+08 | 3.5E+08 | 6.7E+09 | 0.0454 | 0.33 | 0.122 | 1.70 | 0.26 | 2.0 | 5 | NOEMA | 0.6" | 30 |
| 19 | EGS4_24985 | 0.40 | 144 | 2.7E+08 | 3.6E+08 | 6.0E+09 | 0.0414 | 0.30 | 0.083 | 1.30 | 0.51 | 3.8 | 5 | LBT/ NOEMA | 0.85" | 80 |
| 20 | zC_403741 | 0.68 | 103 | 2.1E+08 | 9.7E+07 | -2.4E+09 | 0.0213 | 0.12 | 0.077 | 0.10 | 0.38 | 2.2 | 5 | SINF- H100+H25 0 | 0.35" | 6 |
| 21 | D3a_6397 | 0.57 | 150 | 2.9E+08 | 1.9E+08 | -1.2E+10 | 0.0451 | 0.22 | 0.113 | 1.18 | 0.47 | 1.5 | 5 | SINF-K100 | 0.35" | 9 |
| 22 | EGS_13011166 | 0.55 | 139 | 2.8E+08 | 2.0E+08 | -9.0E+09 | 0.0633 | 0.19 | 0.074 | 0.49 | 0.43 | 2.5 | 5 | LBT/ NOEMA | 0.75" | 15 |
| 23 | GS4_43501 GK_2438 | 0.40 | 123 | 2.7E+08 | 1.8E+08 | -6.9E+09 | 0.0398 | 0.19 | 0.122 | 1.17 | 0.35 | 2.9 | 5 | SINF-H250 / KMOS | 0.55" | 22 |
| 24 | GS4_14152 | 0.23 | 222 | 5.2E+08 | 1.1E+08 | -6.0E+10 | 0.0415 | 0.08 | 0.133 | 0.93 | 0.50 | 2.0 | 5 | SINF-H250 | 0.55" | 12 |
| 25 | K20_ID9- GS4_27404 | 0.30 | 103 | 2.4E+08 | 2.3E+08 | -1.8E+09 | 0.0728 | 0.42 | 0.144 | 1.04 | 0.46 | 1.9 | 4 | SINF-K250 | 0.45" | 6 |
| 26 | zC_405501 | 0.07 | 76 | 2.0E+08 | 2.9E+07 | -1.3E+10 | 0.0255 | 0.15 | 0.162 | 0.59 | 0.39 | 2.4 | 4 | SINF- K100/K250 | 0.3" | 8 |
| 27 | SSA22_MD41 | 0.05 | 54 | 1.1E+08 | 3.2E+08 | 3.3E+10 | 0.0713 | 0.90 | 0.124 | 0.79 | 0.37 | 2.1 | 4 | SINF-K250 | 0.55" | 7 |
| 28 | Q2343- BX_389 | 0.30 | 113 | 3.0E+08 | 3.4E+08 | 6.1E+09 | 0.0621 | 0.40 | 0.132 | 1.53 | 0.29 | 1.5 | 4 | SINF- K100/K250 | 0.45" | 9 |
| 29 | zC_407302 | 0.50 | 85 | 2.5E+08 | 1.4E+08 | -5.2E+09 | 0.0595 | 0.14 | 0.066 | 1.02 | 0.45 | 2.4 | 4 | SINF-K100 | 0.27" | 19 |
| 30 | GS3_24273 | 0.80 | 91 | 2.4E+08 | 1.2E+08 | -1.8E+10 | 0.0714 | 0.25 | 0.112 | 0.07 | 0.18 | 2.4 | 4 | KMOS | 0.57" | 16 |
| 31 | zC_406690 | 0.90 | 91 | 2.7E+08 | 3.2E+07 | -1.5E+10 | 0.0500 | 0.02 | 0.059 | 0.15 | 0.42 | 2.4 | 4 | SINF-K100 | 0.3" | 10 |
| 32 | Q2343- BX_610 | 0.42 | 98 | 2.9E+08 | 3.5E+08 | 4.9E+09 | 0.0555 | 0.29 | 0.134 | 0.86 | 0.46 | 2.0 | 4 | SINF- K100/K250 | 0.6" | 34 |
| 33 | K20_ID7- GS4_29868 | 0.03 | 96 | 2.5E+08 | 4.1E+08 | 3.5E+10 | 0.0849 | 0.57 | 0.108 | 1.30 | 0.21 | 1.7 | 4 | SINF- K100/K250 +KMOS | 0.38" | 33 |
| 34 | K20_ID6- GS3_22466- GS4_33689 | 0.30 | 83 | 2.4E+08 | 1.0E+08 | -1.1E+10 | 0.0288 | 0.30 | 0.147 | 0.11 | 0.30 | 1.8 | 4 | SINF- K100/K250 +KMOS | 0.42" | 26 |
| 35 | zC_400569 | 0.70 | 106 | 3.4E+08 | 3.5E+07 | -1.5E+10 | 0.0361 | 0.05 | 0.097 | 0.14 | 0.36 | 2.0 | 4 | SINF-K100 | 0.35" | 22 |
| 36 | Q2346- BX_482 | 0.02 | 88 | 2.5E+08 | 5.4E+08 | 3.0E+10 | 0.0655 | 0.57 | 0.136 | 0.81 | 0.43 | 2.0 | 4 | SINF- K100/K250 | 0.3" | 18 |
| 37 | COS4_02672 | 0.10 | 91 | 2.6E+08 | 1.2E+08 | -2.3E+10 | 0.0458 | 0.25 | 0.189 | 1.05 | 0.38 | 1.7 | 4 | SINF-K250 +KMOS | 0.5" | 14 |
| 38 | D3a_15504 | 0.30 | 111 | 3.7E+08 | 2.7E+07 | -4.0E+10 | 0.0401 | 0.08 | 0.080 | 0.23 | 0.35 | 2.4 | 4 | SINF- K100/K250 | 0.4" | 47 |
| 39 | D3a_6004 | 0.44 | 142 | 5.2E+08 | 1.0E+08 | -3.6E+10 | 0.0396 | 0.05 | 0.112 | 0.30 | 0.44 | 1.8 | 4 | SINF- K100/K250 | 0.4" | 23 |
| 40 | GS4_37124 | 0.70 | 82 | 3.1E+08 | 3.1E+08 | -3.7E+07 | 0.0375 | 0.25 | 0.069 | 1.17 | 0.34 | 3.5 | 4 | KMOS | 0.43" | 17 |
| 41 | GS4_42930 GK_2363 | 0.50 | 70 | 2.7E+08 | 7.9E+07 | -4.6E+09 | 0.0276 | 0.11 | 0.118 | 0.17 | 0.32 | 1.7 | 4 | SINF-K100 | 0.25" | 20 |